# Reasoning over Taxonomic Change: Exploring Alignments for the *Perelleschus* Use Case

**Nico M. Franz**[1*], **Mingmin Chen**[2], **Shizhuo Yu**[2], **Parisa Kianmajd**[2], **Shawn Bowers**[3], **Bertram Ludäscher**[2]

**1** School of Life Sciences, Arizona State University, Tempe, Arizona, United States of America
**2** Department of Computer Science, University of California at Davis, Davis, California, United States of America
**3** Department of Computer Science, Gonzaga University, Spokane, Washington, United States of America

* E-mail: nico.franz@asu.edu

**Abstract**

Classifications and phylogenetic inferences of organismal groups change in light of new insights. Over time these changes can result in an imperfect tracking of taxonomic perspectives through the re-/use of Code-compliant or informal names. To mitigate these limitations, we introduce a novel approach for aligning taxonomies through the interaction of human experts and logic reasoners. We explore the performance of this approach with the *Perelleschus* use case of Franz & Cardona-Duque (2013). The use case includes six taxonomies published from 1936 to 2013, 54 taxonomic concepts (i.e., circumscriptions of names individuated according to their respective source publications), and 75 expert-asserted Region Connection Calculus articulations (e.g., congruence, proper inclusion, overlap, or exclusion). An Open Source reasoning toolkit is used to analyze 13 paired *Perelleschus* taxonomy alignments under heterogeneous constraints and interpretations. The reasoning workflow optimizes the logical consistency and expressiveness of the input and infers the set of maximally informative relations among the entailed taxonomic concepts. The latter are then used to produce merge visualizations that represent all congruent and non-congruent taxonomic elements among the aligned input trees. In this small use case with 6-53 input concepts per alignment, the information gained through the reasoning process is on average one order of magnitude greater than in the input. The approach offers scalable solutions for tracking provenance among succeeding taxonomic perspectives that may have differential biases in naming conventions, phylogenetic resolution, ingroup and outgroup sampling, or ostensive (member-referencing) versus intensional (property-referencing) concepts and articulations.

**Introduction**

The present contribution is a companion paper to [1] and offers a novel, use case-centered illustration of *aligning* multiple succeeding taxonomies through the interaction of human experts[1] and logic reasoners. It is of a technical, detail-focused nature and most immediately directed at biodiversity scientists who wish to integrate taxonomically non-congruent classifications and phylogenies. In spite of the technical presentation, the impacts of logically representing stability and change across taxonomies may be wide-ranging. Our broader intention is to promote the *taxonomic concept approach* [2–4] as a feasible solution to the challenge of provenance tracking in cases were name/circumscription relationships change across taxonomies as an outcome of scientific advancement. The central objective of this approach is to *represent* the change, not to judge the correctness of the original and revised taxonomies. Hence, in aligning multiple input classifications or phylogenies, we ask not: "which names or circumscriptions are valid?" Instead we ask: "how can we logically represent, and thus perform reliable inferences over, the similarities and differences between multiple, independently published taxonomic perspectives?" We show here that this is feasible with an unprecedented degree of resolution.

The publication [1] preceding our analysis entails a phylogenetic revision of the weevil genus *Perelleschus* O'Brien & Wibmer sec. Franz & Cardona-Duque (2013). In that revision a new arrangement of the ten entailed species (concepts) was presented, along with five relevant prior classifications and phylogenies published from 1936 to 2006. In all, these six taxonomies entailed 54 *taxonomic concepts* [2–4], which in turn were aligned by the authors using 75 *articulations* [1,4–7]. Jointly the concepts, hierarchies, and articulations provide the baseline input for the logic-facilitated alignments. To motivate our analysis, we first review the background and challenges related to task of reasoning over taxonomic change [8].

## The Challenge of Tracking Context-Contingent Name/Circumscription Relationships

Why might reasoning over taxonomic change be desirable? In addressing this question, we may take as well established that classifications and phylogenies of organismal groups change in light of new insights [4,9–10]. Over time these changes can result in an imperfect semantic tracking of taxonomic perspectives through varying combinations of valid names and synonyms [1–3,11–12]. It is generally recognized that the Linnaean naming system, while serviceable in many regards [13–15], is not designed to represent all kinds of taxonomic content change; i.e., to fully track taxonomic *provenance* [16–18]. This circumstance poses different data integration challenges depending on the agents – humans versus computers – and space/time dimensions involved in name-based data transmission.

The traditional way to transmit taxonomically annotated content is through human-to-human communication, including communication via publications. In these situations the inability of the Linnaean names to represent any and all taxonomic similarities and differences is often mitigated by the communicators' shared ability to infer the relevant *context* in which name-based information is exchanged [19]. For instance, if two biologists convened today to discuss the relationships among families of beetles, most likely both would have in mind taxonomic definitions for those families that are largely contemporary and therefore also congruent. As an approximate rule, then, inter-human communication via taxonomic names succeeds to the extent that the speakers' reciprocal assumptions about the contexts of name/circumscription relationships coincide.

The mitigating effects of human comprehension of implied contexts tend to break down over greater space-time dimensions, with concomitant losses of precision in communication.

---

[1] For the purpose of this paper, "expert" is broadly defined as: a person (or persons) with (some) knowledge of certain taxonomic groups and motivation to assert an initial set of concept articulations.

Typically only experts have the ability to interpret names for higher taxa of beetles as circumscribed in the late 1700s [20] and align their meanings confidently with current circumscriptions [21] – even though many names have remained valid throughout the centuries-long time span. Consequently, the reliable long-term use of taxonomic names as proxies for evolving circumscriptions of taxonomic entities requires a continuous process of tracking name/circumscription relationships [4,9–10]. The process is carried out primarily by expert authors whose novel insights are passed on through (e.g.) revised classifications and phylogenies to other expert and non-expert user communities [22–24]. However, the cultural practice of communicating and updating human speakers on the taxonomic content of names is not readily transferable to the realm of logic representation.

The challenge we wish to address is systemic: how can we represent complex taxonomic changes computationally based on type-fixated name equivalences, and do so without compromising the valuable services provided by the Linnaean naming system to facilitate human communication about perceived taxa? Tracking imperfectly co-evolving name/circumscription relationships is challenging enough for humans who bring considerable cognitive abilities to bear to this endeavor. The challenge becomes insurmountable, however, in the strict logics framework that sustains the Semantic Web [3,7,12,25–26,27–28]. While taxonomic names are necessary to facilitate integration of biological information [14], they are not sufficient for this task [4–5,8]. This is so because the traditional, type-centered conventions under which names and nomenclatural relationships are created [29–31] render these identifiers taxonomically *underspecified* [1–4,11–12]. Whereas humans can discern context based on shared communication norms, computers are more limited in understanding what names such as "*Perelleschus*" mean 'under relevant conditions' [32]. Thus improved annotation practices are needed to achieve reliable name/circumscription representation in the computational realm [7,33–34].

Additional specification of context, and hence improved identifier resolution, may be provided through the taxonomic concept approach [2,4,9]. Under this approach, the names and circumscriptions are individuated *according to* ("sec.") the corresponding authors or references. This is reflected in the convention to label concepts as (e.g.) *Perelleschus* Wibmer & O'Brien 1986 (name) *sec.* Franz & Cardona-Duque 2013 (source), thereby identifying the latter authors' (2013) taxonomic redefinition of the former authors' (1986) name and circumscription [1]. In a subsequent step, each concept can be aligned via articulations that specify taxonomic congruence or non-congruence at a finer semantic scale than the name-based system [6,9,35].

In the workflow detailed below (figures 1 and 2), an initial set of articulations is provided by an expert. The alignment is then optimized for consistency (absence of conflict) and expressiveness (absence of ambiguity) using the capabilities of logic reasoners. The final product is a *merge taxonomy* that represents the concept-level similarities and differences among the aligned input trees; i.e., a precise semantic map showing how to navigate among the taxonomic concepts entailed in each source tree [8].

Reasoning over Taxonomic Change – Antecedents

Logic-facilitated reasoning over taxonomic change is not computationally trivial and remains in a developmental stage. A brief history of this special case of ontology matching [36] under taxonomic constraints is offered here to orient the reader.

The use of taxonomic concepts in biodiversity databases dates back to the early 1990s [2,37–38]. The use of concept-to-concept articulations was pioneered in [9,35,39]. The approach was developed more formally in [5,40–41] through the notion of a potential taxon *graph* that can transmit information linked to succeeding concepts under logic constraints. The graph edges and alignments are grounded in *Region Connection Calculus* (RCC-5) relations, herein also called articulations [42]. RCC-5 entails five basic set theory relationships used to characterize the referential extensions of two taxonomic concepts; viz. *congruence* (==), *proper inclusion* (>), *inverse proper inclusion* (<), *overlap* (><), and *exclusion* (|) [4,6,9]. Ambiguity in an expert's assessment of input articulations can be expressed with the disjunction *or* (e.g., congruence *or* proper inclusion). Combinations of basic and disjunct relationships yield a lattice of 32 possible articulations; beginning with an empty set (∅) at the bottom and ending with a maximally ambiguous set (== or > or < or >< or |) at the top [40,43–44].

Thau and co-authors [43–47] formalized taxonomy alignment in First-Order Logic (FOL) [48]. Their representation was designed to ingest two input taxonomies ($T_1,T_2$ – each assembled via *is_a* edges), an initial set of articulations (A), and additional constraints (C). Constraints relevant to taxonomic reasoning include: (1) *non-emptiness* – a given taxonomic concept has minimally one representing instance; (2) *sibling disjointness* – two given child concepts of a parent concept are exclusive of each other; and (3) *coverage* – a given parent concept is completely circumscribed by the union of its children. Each of these constraints can be relaxed to yield alternative alignment outcomes. The formalizations were first implemented in the prototype software CleanTax [49–50].

**Aligning Real-Life Classifications and Phylogenies – Challenges and Solutions**

Here we explore new solutions for taxonomy alignment under RCC-5, based on the development of the Euler/X toolkit [8,51–54] and its application to the *Perelleschus* use case [1]. Euler/X is an Open Source software toolkit that supersedes CleanTax, incorporating a diverse suite of reasoning services, knowledge products, and improved performance and usability. The toolkit consists of a set of Python programming scripts, multiple logic reasoners [50,55–56], and a tree graph visualization system [57]. Euler/X provides an interactive workflow (figures 1 and 2) guiding users through the process of computing a well-specified alignment (maximal consistency, minimal ambiguity) that corresponds to the input articulations and produces a merge visualization. The toolkit allows encoding the input constraints ($T_1,T_2,A,C$) as a Stable Model Semantics (SMS) problem [58], and utilizes powerful Answer Set Programming (ASP) reasoners [56,59–60] to achieve alignments in efficient time [51–52,54].

At present the Euler/X toolkit is suited to analyze pairwise, small- to medium-scale real-life use cases of 100-500 concepts per input taxonomy [51]. Step-wise explorations of such use cases are critical to a wider adoption, because this demonstrates feasibility and illustrates how an expert's input articulations and related assumptions about taxonomy alignment *interact* with the logic of reasoners to produce the desired outcomes. In general, the alignment workflow should leverage the strengths of both (1) the expert to assert non-/congruence among sets of concepts, and (2) the reasoners to perform myriads of contingent logic proofs to infer complete and consistent alignments [44–46]. In the next sections we introduce three categories of challenges that the expert/toolkit interaction must resolve.

Heterogeneous Taxonomic Coverage Constraints

Classifications and phylogenies are representations of heterogeneous systematic analyses. The representations often have taxonomic, geographic, or rank-specific constraints [7]. Processing these constraints is critical to the generation of informative alignments. For instance, regional catalogues tend to omit non-focal taxon diversity and may not reflect the latest phylogenetic insights [61]. In contrast, higher-level classification synopses may wholly omit genus- and/or species-level concepts [21]. Phylogenetic analyses often undersample the lowest taxonomic levels [62]. They typically include both ingroup and outgroup entities, yet the latter may be less focal to the alignment process (instead introducing 'noise'). Phylogenies also make use of informal ranks or leave certain lineages unnamed [63]. Other special constraints include monotypic parent concepts whose children have different nomenclatural ranks but congruent taxonomic extensions.

Alignments of disparate input trees are possible but outcomes will vary depending on whether the *coverage* constraint is globally enforced or locally relaxed [6–8]. Coverage is also relevant to differential *readings* of articulations, focusing either on strictly member-based or property-based components of the respective concepts, or both [1,4,6].

Consistency, Exhaustiveness, and Expressiveness

Given two taxonomies ($T_1,T_2$) and additional constraints (C), an expert's set of input articulations (A) can yield three reasoning outcomes (figure 1). (1) The set of input articulations is logically consistent, well specified, and thus yields a single alignment or *possible world* that satisfies all constraints. This is the intended outcome [8], however due to the difficulty of asserting articulations it is not always the initial result. (2) The set of input articulations is logically consistent but retains sufficient ambiguity to yield multiple possible world alignments [51]. (3) The input constraints are *not* logically consistent. This means that unless one or more combinations of these constraints are 'adjusted' to agree, there is no stable model and hence no alignment. Additional actions are needed to diagnose and repair the constraint(s) causing the inconsistency.

The three outcomes inform optimization criteria for the toolkit workflow (figure 1). Accordingly, the demands on the final set of *output* articulations are: (1) *Consistency* – the articulations cannot logically contradict each other. For instance, two concepts A and B cannot reciprocally include each other (A > B and B > A). The reasoning process must identify instances of inconsistency in the input and lead to their resolution (for further details see [53–54]). (2) *Exhaustiveness* – users cannot be expected to provide articulations for *all* (pairwise) combinations of concepts represented in the input taxonomies. Instead they will provide a relevant *subset* of articulations [6], relying on the reasoning process to infer the logically implied articulations that amount to an exhaustive alignment. Thau and co-authors [44–46] use the term *maximally informative relations* (MIR) to refer to that set which includes for all paired concepts the 'true' articulation, i.e., the one based on which all other true relations in the 32-relations lattice can be deduced. An exhaustive alignment is obtained when the expert's input articulations are expanded by the reasoning process to yield the set of MIR. (3) *Expressiveness* – alignments that are consistent and exhaustive may nevertheless retain inherent ambiguities, leading to numerous merge taxonomies [51]. For instance, the number of possible worlds increases dramatically if multiple disjoint articulations are represented in the input. In some instances the entire set of possible world solutions will become the final outcome, and may be summarized through tree consensus methods [64]. In other cases the user can *reduce* this set by working through an interactive decision tree process [65]. This process can lend additional specification to the set of input articulations and therefore make the outcome more expressive. The toolkit iteratively identifies questions of the type "are these two concepts congruent or are they overlapping?" and then eliminates possible worlds that are no longer implied by the answer.

Merge Concept Representation

Articulations of congruence (==), inclusion (>,<), or exclusion (|) generate merge taxonomies that preserve an arrangement of nested, properly inclusive relationships among the input concepts. This means that the input concept *labels* are adequate for referring to each merge region [66]. In contrast, *overlapping* articulations generate novel concept partitions (figure 2). In the simplest case, two overlapping input concepts A and B will create three merge regions, as follows: the region that is unique to A (A, *not* b; or written as A\B), the region that is unique to B (*not* a, B; or B\A), and the region that constitutes the overlap (AB; or A*B). We refer to these novel partitions as *merge concepts* (see also [67]). Representing merge concepts provides a better understanding of the extent of concept overlap.

**Materials and Methods**[2]

The *Perelleschus* use case [1] includes six taxonomies (figure 3) that represent classificatory changes related to the species-level concept *Elleschus carludovicae* Günther sec. Günther (1936) [68]. The history spans from the concept's origin in 1936 to the most recent revision in 2013. Changes undertaken in this interval include: the generic reassignment to *Perelleschus* Voss sec. Voss (1954) [69] and nomenclatural validation of *Perelleschus* Wibmer & O'Brien sec. Wibmer & O'Brien (1986) [61]; species-level concept additions [1,69–70], and phylogenetic redefinitions of the genus-level concept [1,62,70].

Diagnostic details on this Neotropical lineage of Acalyptini Thomson [non-focal] flower weevils (Coleoptera [non-focal]: Curculionidae [non-focal]) are provided in [1]. In reference to the aforementioned representation challenges, we note that three of the taxonomies selected for alignment are traditional, ranked classifications [61,68–69]. The remaining three taxonomies are phylogenies that also include informally named, synapomorphy-carrying clades [1,62,70]. One of these [62] represents an exemplar study that is undersampled at the lowest taxonomic level [71].

Our approach is to take the 75 articulations published in [1] and linking 54 concepts across six taxonomies *as is*. Thus in several, (we think) generally instructive instances the original input is *not* well specified, instead requiring diagnosis and removal of inconsistent articulations or further specification. We consider original and revised input articulations to distinguish the different stages in the alignment process.

Input Configuration, Workflow Execution, and Concept Labeling Conventions

We analyze the following pairwise alignments (figure 3): (1) Günther (1936) [68] and Voss (1954) [69]; (2) Voss (1954) [69] and Wibmer & O'Brien (1986) [61]; (3) Wibmer & O'Brien (1986) [61] and Franz & O'Brien (2001) [70]; (4) Franz & O'Brien (2001) [70] and Franz (2006) [62]; (5) Franz (2006) [62] and Franz & Cardona-Duque (2013) [1]; and (6) Franz & O'Brien (2001) [70] and Franz & Cardona-Duque (2013) [1]. The first five of these provide a chronological chain spanning the entire 1936-2013 time period (see also [8]). The sixth alignment is significant because it concerns two highly comparable phylogenetic revisions.

---

[2] In the subsequent sections we adopt the convention of [1], where: (1) the name sec. author annotation is used whenever a specific circumscription of a name is intended; (2) just the name is used to refer to the cumulative legacy of circumscriptions associated with that name; and (3) the term [non-focal] is appended to a name when specifying its meaning is outside of the present scope. We spell out the taxonomies' authors for clarity where appropriate.

Instructions for the installation and command-line operation of the Euler/X toolkit are specified in [8,51–52]). An overview of the toolkit's options is provided through the `"euler --help"` function. The input (.txt) files corresponding to figures 4-16 were assembled manually based on the information in tables 2-4 in [1] and workflow-diagnosed revisions. They are appended to this paper in the Supporting Information S1.

We list the more recently published taxonomy (i.e., $T_2$ in each alignment) and concept(s) first in the input files and when denoting the input articulations (see also figure 2). The toolkit run commands are provided as annotations (#...) at the beginning of the input file. For simplicity's sake we address the two main options as (1) containment with overlap analysis or graph (command: `-e mnpw --rcgo`; figure 2E) and (2) merge concept analysis or graph (command: `-e mncb`; figure 2F).

To economize space in the text and merge visualizations, the concept labels of [1] are abbreviated as shown in figure 3; e.g. *Perelleschus carludovicae* (Günther) sec. Wibmer & O'Brien (1986) is abbreviated as "1986.Pcarlud" (figure 3C). Abbreviations for above species-level concepts are capitalized, as in "2006.PER" for *Perelleschus* Wibmer & O'Brien sec. Franz (2006) (figure 3E). Binomials include both upper- and lowercase letters, as in 2001.PHYsubcin for *Phyllotrox subcinctus* (Voss) sec. Franz & O'Brien (2001). To abbreviate informally assigned clade names we introduce an underscore ("_"), as in "2001.Peve_Pvar" for the *Perelleschus evelynae-Perelleschus variabilis* clade sec. Franz & O'Brien (2001). These abbreviated concept labels are also used in the toolkit data input files.

**Results**

**Alignment 1 – Voss (1954) and Günther (1936)**

Alternative Interpretations of Higher-Level Input Articulations

The alignment of the Voss (1954) and Günther (1936) classifications (figures 3-8) presents a microcosm of more generally occurring challenges for this alignment approach. Accordingly, the toolkit's abilities to represent alternative perspectives and levels of resolution will be treated in detail.[3] We anticipate that future results derived from the toolkit will be more streamlined and guided by efficient user interfaces. We first discuss multiple interpretations of this alignment, then proceed to alignment realizations, and lastly address the issue of identifying the proper levels of input specification.

Much of the ambiguity originates with Günther (1936: 190-191) who places the newly named species-level concept *E. carludovicae* sec. Günther (1936) into a genus-level entity *Elleschus* sec. Günther (1936). The latter carries a previously coined name whose circumscription is in effect *expanded* by inclusion of the new child concept. But the author does not offer a detailed genus-level revision, instead referring to a preceding monograph [72] and noting that his species-level concept "stands passably close to the species of the genus *Elleschus*" [1]. These additional, implied species-level concepts are *not* mentioned *in name* by Günther (1936); however Lacordaire (1863: 606) lists two such concepts of which the 1936 author was presumably aware.

---

[3] For ease of reading, we recount the creation of input taxonomies and articulations in present tense.

Voss (1954) creates a new genus-level concept *Perelleschus* sec. Voss (1954) which includes *P. carludovicae* sec. Voss (1954) in addition to two newly named and circumscribed species-level concepts. At the species level, then, the articulations are readily apparent (figures 4-8); viz. 1954.Pcarlud == 1936.ELLcarlud, 1954.Prectir | 1936.ELLcarlud, and 1954.Psubcin | 1936.Ecarlud.

Ambiguity is introduced when the genus-level concepts are aligned. Congruence (1954.PER == 1936.ELL) is not a viable articulation because Voss (1954) explicitly separates the referential extension of *Perelleschus* sec. Voss (1954) from those of preceding concepts linked to the name *Elleschus*. Exclusion (1954.PER | 1936.ELL) is also not adequate due to the inclusion of the congruent child-level concepts 1954.Pcarlud and 1936.ELLcarlud, respectively. The remaining three articulations – i.e., 1954.PER > or >< or < 1936.ELL – are all potentially valid [1]. Selecting any of these articulations reflects alternative interpretations of explicit or implicit information inherent in the two source classifications. We examine each of the three readings.

***Perelleschus* sec. Voss (1954) properly includes *Elleschus* sec. Günther (1936).** We may call this an *explicit* and *strictly ostensive* reading. Ostension is understood here as the practice of representing the meaning of a class term by *pointing to* one or more of its referent *members* [73]. Intension, in turn, is the practice of defining the referential extension of a class term by *asserting the properties* that referent entities must exhibit. Taxonomies often make use of intensional/ostensive *hybrid* definitions for their constituent concepts [4,6–7,74–75]. We use the abbreviations [INT] and [OST] of [6] to indicate whether an intensional or ostensive reading of an articulation is applied. [INT/OST] means that either interpretation is permissible. Modulation of the *coverage* constraint is critical in this context, as demonstrated below.

In an explicit and strictly ostensive sense, parent concept 1954.PER has three child concepts 1954.Pcarlud, 1954.Prectir, and 1954.Psubcin; whereas parent concept 1936.ELL has one child concept 1936.ELLcarlud; and 1954.Pcarlud == 1936.ELLcarlud; hence 1954.PER > 1936.ELL [OST].

***Perelleschus* sec. Voss (1954) overlaps with *Elleschus* sec. Günther (1936).** We may call this an *implicit* reading under either intensional or ostensive conditions. Accordingly, parent concept 1954.PER includes two child concepts 1954.Prectir and 1954.Psubcin that have no congruent entities subsumed under parent concept 1936.ELL. Conversely, parent concept 1936.ELL includes one or more implicit child concepts (likely matching those of Lacordaire [1863]) that have no match under parent concept 1954.PER. The only reciprocally entailed and congruent child concept pair is 1954.Pcarlud == 1936.ELLcarlud; hence 1954.PER >< 1936.ELL [INT/OST].

An ostensive reading of this overlap (1954.PER >< 1936.ELL [OST]) effectively 'elevates' the unmentioned child concepts of *Elleschus* sec. Günther (1936) to an explicit status, which the logic representation must then account for. Alternatively, an intensional reading of the overlap (1954.PER >< 1936.ELL [INT]) asserts the following: had the preceding author Günther (1936) examined voucher material pertaining to the species-level concepts 1954.Prectir and 1954.Psubcin as recognized by Voss (1954), then he would *not* have subsumed them under *Elleschus* sec. Günther (1936), presumably because of an insufficient match of genus-level characters. This is a negation of a counterfactual assertion – there is no evidence that Günther (1936) examined voucher material pertaining to Voss' (1954) two new species-level concepts, or even expressed an assessment as to their taxonomic identity. However, because intensional definitions have predictive powers [7], this forward-looking interpretation is allowable in principle.

***Perelleschus* sec. Voss (1954) is properly included in *Elleschus* sec. Günther (1936).** We may call this an *implicit* and *intensional* reading. It constitutes the affirmation of the counterfactual assertion: had the preceding author Günther (1936) examined material of the species-level concepts 1954.Prectir and 1954.Psubcin as recognized by Voss (1954), then he *would* have subsumed them under *Elleschus* sec. Günther (1936), presumably because of a sufficient match of genus-level characters. In that sense, 1954.PER < 1936.ELL [INT].

Alignment Realizations

Each of the three interpretations can represent (figures 4-6). The original input articulations of [1] are shown in figure 4; viz. 1954.PER > or >< or < 1936.ELL. Under both the containment with overlap and merge concept analyses, this input yields a single, unambiguously resolved possible world alignment in which 1954.PER > 1936.ELL is the consitent articulation for the genus-level relationship (figure 4). The merge shows *Elleschus* sec. Günther (1936) and *E. carludovicae* sec. Günther (1936) as properly included with *Perelleschus* sec. Voss (1954) *and* congruent with *P. carludovicae* sec. Voss (1954). The output MIR include eight articulations, of which four are immediately deducible and four are inferred by the reasoner (figure 4).

Using just the 1954.PER > 1936.ELL articulation in the input returns the same outcome. The three unambiguous species-level articulations, together with the standard constraints, allow *only* the 1954.PER > 1936.ELL [OST] interpretation to remain consistent. Thus by default, in the reasoner will constrain underspecified articulations among parent concepts *based strictly on ostension* to the corresponding, unambiguously articulated child concepts. Because *Elleschus* sec. Günther (1936) has a single child concept that fully defines its extension in the logic framework, no other outcomes are consistent.

In order to express an overlapping genus-level articulation 1954.PER >< 1936.ELL [INT/OST], the input alignment must be *modulated* as follows (compare figures 4 and 5): (1) restrict the input articulation to 1954.PER >< 1936.ELL; (2) introduce an 'implied concept' 1936.ELL_IC as an additional child of concept 1936.ELL; and (3) add the input articulation 1954.PER | 1936.ELL_IC. Jointly these modifications have the effect of allowing other (implied) children to be accounted for under *Elleschus* sec. Günther (1936), thereby logically representing the author's obscure reference to Lacordaire (1863) and the therein recognized species-level concepts. The 1954.PER | 1936.ELL_IC articulation is needed to specify that no child concepts implied in Günther (1936) intersect with the newly recognized child concepts in Voss (1954). The output MIR include seven immediately deducible and five reasoner-inferred articulations (figure 5).

The containment with overlap graph (figure 5A) illustrates that *Perelleschus* sec. Voss (1954) and *Elleschus* sec. Günther (1936) intersect *exactly* though shared inclusion of the congruent species-level concepts 1954.Pcarlud == 1936.ELLcarlud. The remaining three child-level concepts are not shared among the parent concepts. In the merge concept analysis (figure 5B), the shared and unique regions resulting from the genus concept-level overlap are resolved separately. This representation yields additional merge concept labels for two of the input concept clusters; i.e. viz. 1936.ELL*1954.PER – the region that constitutes the overlap – and 1936.ELL\1954.PER (where the "\" means *not*) – the region (also labeled 1936.ELL_IC) which is unique to Günther's (1936) genus-level concept. Lastly, through combination of input concepts 1954.Prectir and 1954.Psubcin, the concept merge analysis creates a new merge concept 1954.PER\1936.ELL that is unique to Voss (1954). Articulations to and from this merge concept are shown are newly inferred (red color; figure 5B).

The third, intensional representation 1954.PER < 1936.ELL [INT] similarly requires addition of an implied child concept (1936.ELL_IC) and specification of the genus-level articulation 1954.PER < 1936.ELL (figure 6). However, if the implied concept 1936.ELL_IC is allowed to include regions subsumed under 1954.PER that are not *P. carludovicae* sec. Voss (1954), then the additional input articulation 1954.PER | 1936.ELL_IC (see figure 5) is *not* required. Based on the modulated input constraints, a single possible world merge is obtained (figure 6). The output MIR include six immediately deducible and six reasoner-inferred articulations. The reasoner infers that both species-level concepts 1954.Prectir and 1954.Psubcin of Voss (1954) are properly included within the genus-level concept *Elleschus* sec. Günther (1936). The containment with overlap graph (figure 6A) aligns these children as jointly subsumed under the implied concept 1936.ELL_IC, which corresponds to the affirmative counterfactual assertion that Günther (1936) would have acted accordingly. Because the input articulations stipulate that 1954.PER < 1936.ELL, the analysis infers an overlap (><) among concepts 1954.PER and 1936.ELL_IC. This amounts to the assertion that concept 1936.ELL_IC, and by extension concept 1936.ELL, include regions *other than those* subsumed under concept 1954.PER. In order words, the shared region constituting 1954.PER >< 1936.ELL_IC corresponds *exactly* to the union of concepts 1954.Prectir and 1954.Psubcin. The merge concept graph (figure 6B) differentiates this region into two subregions – 1936.ELL_IC\1954.PER and 1936.ELL_IC*1954.PER – that account for the shared and non-shared elements of the implied concept.

We may summarize the three alignment realizations as follows. By default the reasoner assesses congruence and non-congruence among higher-level concepts directly based on ostension to their respective child concepts (figure 4). This representation, rooted in the coverage assumption [43–45], can be modulated through introduction of implied concepts *and* stipulation of specific higher-level articulations that yield alternative alignments. The latter, in turn, can be interpreted – either in additional (figure 5) or exclusively (figure 6) – as intensional alignment representations.[4] The alternative visualizations resolve only input (figures 5A and 6A) or also merge concept regions (figures 5B and 6B).

Consistency and Expressiveness

The option to assert alternative intensional or ostensive readings for higher-level articulations has implications for the toolkit's usability. To minimize human effort in specifying the input, experts should have an understanding of the proper levels of *sufficiency* (Table 1); i.e., how to obtain the minimal, well-specified input alignment (figure 1). The three sets of input articulations depicted in figures 4-6 all fulfill the criterion of sufficiency by yielding unique merges. Indeed, the input for the 1954.PER > 1936.ELL [OST] alignment (figure 4) is *overspecified;* eliminating the genus-level articulation (1954.PER > or >< or < 1936.ELL) produces an identical outcome. If instead we provide as input the unambiguous genus-level articulations 1954.PER >< 1936.ELL or 1954.PER < 1936.ELL, yet without adding an implied child 1936.ELL_IC, then the reasoner infers that the input is inconsistent. The toolkit launches an inconsistency explanation module [53–54] and suggests one available repair option; i.e., the removal of the genus-level articulation. We may provisionally infer that specifying a maximally informative (non-disjoint) set of input articulations *among the lowest-level concepts* is sufficient to yield a well-specified alignment *if* coverage (by ostension) holds throughout each input taxonomy. This 'rule' indicates that the genus-level articulation 1954.PER > 1936.ELL is logically redundant under the ostensive reading.

---

[4] The input articulations for the 1954.PER >< 1935.ELL alignment could be further modified to assert that *Elleschus* sec. Günther (1936) may include only one (say, concept 1954.Prectir) of the two species-level concepts newly described in Voss (1954) but not the other (concept 1954.Psubcin, accordingly).

The 1954.PER >< 1936.ELL alignment (figure 5), in turn, requires more specification. The cause of this is the additional (implied) concept 1936.ELL_IC, assigned to Günther (1936), whose articulations to concepts in Voss (1954) would otherwise remain underspecified. Omitting the 1954.PER | 1936.ELL_IC articulation either individually (figure 7), or jointly with the 1954.PER >< 1936.ELL articulation (figure 8), yields 8 and 17 possible world merges, respectively. In the former case (figure 7), the output MIR include five inferred, disjoint articulations. The seven additional possible world alignments (figures 7B–H) can be characterized as follows: two alignments with 1954.Prectir < 1936.ELL_IC (7E, 7F), two alignments with 1954.Psubcin < 1936.ELL_IC (7G, 7H), one alignment with 1954.Prectir >< 1936.ELL_IC (7B), one alignment with 1954.Psubcin >< 1936.ELL_IC (7D), and one alignment with 1954.Prectir >< 1936.ELL_IC *and* 1954.Psubcin >< 1936.ELL_IC (7C). Thus in absence of the input articulation 1954.PER | 1936.ELL_IC, the implied child concept can variously include or overlap with the species-level concepts 1954.Prectir and 1954.Psubcin of Voss (1954), rendering the output ambiguous. This concept behaves like a 'floater'. In the latter, even less specified case (figure 8), the output MIR have six inferred, disjoint articulations. As a consequence, the genus-level concepts 1954.PER and 1936.ELL can articulate via any relationship except exclusion, producing nine additional possible worlds. In seven of these, 1954.PER > 1936.ELL, and only one merge includes 1954.PER < 1936.ELL, reflecting the intensional reading (figure 8C).

**Alignment 2 – Voss (1954) and Wibmer & O'Brien (1986)**

The alignment of the Voss (1954) and Wibmer & O'Brien (1986) classifications is straightforward (figure 9). Changes made in the latter are grounded in nomenclatural validity and imply no taxonomically incongruent views [1]. Each concept of the earlier treatment has a congruent match in the later treatment. Of the 16 output MIR, 12 are immediately deducible (Table 1). Removal of the logically redundant genus-level articulation 7 == 3 yields the identical result. Alternative intensional/ostensive interpretations or containment are not under consideration in this alignment. The containment with overlap and merge concepts graphs are the same.

**Alignment 3 – Wibmer & O'Brien (1986) and Franz & O'Brien (2001)**

The alignment of the Wibmer & O'Brien (1986) and Franz & O'Brien (2001) perspectives is the first in this sequence to merge a traditional, ranked classification with a more resolved phylogeny that contains informal clade names and concepts (figure 3). Three kinds of taxonomic changes are undertaken in Franz & O'Brien (2001) in relation to the preceding classification: (1) *Perelleschus* sec. Franz & O'Brien (2001) is redefined through a set of perceived synapomorphic properties. The redefinition necessitates the removal of *Perelleschus subcinctus* sec. Voss (1954) from the revised genus concept, transfer to *Phyllotrox* sec. Franz & O'Brien (2001), and renaming (new combination) to *Phyllotrox subcinctus* sec. Franz & O'Brien (2001). (2) Six new species-level concepts are added to *Perelleschus* sec. Franz & O'Brien (2001). (3) The eight recognized members of the genus-level are arranged into five informally named clade concepts below the genus level. Hence the alignment must represent changes commonly encountered when comparing traditional classifications and phylogenetic revisions; including the addition of new terminals and new or redefined clades.

Of the seven input articulations provided in [1], one articulation (2001.PER == or > (1986.PER – 1986.Psubcin)) cannot be represented in the input because it involves subtracting a child concept (1986.Psubcin) from its parent concept (1986.PER) – an operation that not yet supported. The remaining six input articulations (figure 10) are not further differentiated into intensional or ostensive components [1]. Nevertheless they are jointly *inconsistent* when supplied to the reasoner. The toolkit identifies two repair options: (1) removal of the genus-level articulation 2001.PHY >< 1986.PER, or (2) removal of the species-level articulation 2001.PHYscubcin == 1986.Psubcin. Each repair path yields a single, consistent alignment, but only the former option is deemed appropriate (figure 10).

In analogy to the challenges of unambiguously representing *Elleschus* sec. Günther (1936) in alignment 1, the source of the inconsistency for alignment 3 is the differential interpretation of coverage for the genus concept *Phyllotrox* sec. Franz & O'Brien (2001), which receives only peripheral treatment in the latter revision. While the authors indicate (2001: 274) that "*Phyllotrox* […] now has 43 species", these implied child concepts are not actually listed in the 2001 perspective, and therefore are not provided to the reasoner. An overlapping 2001.PHY >< 1986.PER articulation *may* be rendered consistent *if* these implied child concepts are represented. However, because the original input of [1] only specifies 2001.PHYsubcin as a child of 2001.PHY, the reasoner infers in the repair that 1986.PER > 2001.PHY. This is not an adequate interpretation of Franz & O'Brien's (2001) phylogenetic revision.

Instead of removing the 2001.PHY >< 1986.PER articulation from the input, it is more appropriate to introduce an implied child 2001.PHY_IC of *Phyllotrox* sec. Franz & O'Brien (2001). This input modulation produces a single alignment, based on 12 immediately deducible and 60 inferred articulations in the output MIR (figure 10; see also Supporting Information S2). The alignment shows the intended genus-level concept relationships 2001.PHY >< 1986.PER and 2001.PER >< 1986.PER, logically grounded in the shared inclusion of congruent species-level concept pairs 2001.Prectir/1986.Prectir and 2001.PHYsubcin/1986.Psubcin, and in the reciprocal exclusion of other concepts (2001.PevePsul and 2001.PHY_IC, respectively). In the containment with overlap analysis (figure 10A), four additional overlapping articulations of the informal clade concepts 2001.Peve_Psul, 2001.Pbiv_Psul, 2001.Ppub_Psul, and 2001.Pcar_Psul with *Perelleschus* sec. Wibmer & O'Brien (1986) are inferred. These clade concepts cumulatively entail the six new species-level concepts as well as the species-level concept pair 2001.Pcarlud/1986.Pcarlud, while excluding the congruent species concepts 2001.PHYsubcin/1986.Psubcin assigned to *Phyllotrox* sec. Franz & O'Brien (2001). The merge concept analysis (figure 10B) resolves each of the six overlapping articulations into narrower Euler regions. Four of these merge concepts 'cascade up' the internal nodes of *Perelleschus* sec. Franz & O'Brien (2001) through newly inferred inclusion relationships, reflecting the addition of six species-level concepts in comparison to the identically named *Perelleschus* sec. Wibmer & O'Brien's (1986).

After addition of the implied child concept 2001.PHY_IC, the input articulations in figure 10 are consistent and overspecified. The three species-level articulations are sufficient to generate the single merge (Table 1).

**Alignment 4 – Franz & O'Brien (2001) and Franz (2006)**

The alignment of the Franz & O'Brien (2001) and Franz (2006) perspectives compares two phylogenetic trees. However, the latter represents an exemplar analysis in which the species level is undersampled [71]. Based on the representation of this phylogeny in [1] (see also figure 3), *Phyllotrox* sec. Franz (2006) is 'childless' and *Perelleschus* sec. Franz (2006) includes only three species-level concepts and one internal clade concept.

In processing the unaltered input of [1], three problems are apparent (figures 11 and 12). First, the top-level articulation 2006.PHYLLO >< 2001.DER provided in [1] is valid only in an intensional sense.[5] This articulation is identified as inconsistent by the reasoner because there is no child of 2006.PHYLLO that *lacks* a congruent match within the tribal-level concept 2001.DER. Hence addition of the articulation 2006.PHYLLO < 2001.DER – resulting in a disjoint articulation 2006.PHYLLO < *or* >< 2001.DER – is needed to represent a wider range of interpretations (including and ostensive reading). Second, the overlapping articulation 2006.PHY >< 2001.PHY is inferred as inconsistent due to the discrepancy of explicitly and implicitly included children of the non-focal genus concept pair *Phyllotrox* sec. Franz (2006) (with no child concept) and *Phyllotrox* sec. Franz & O'Brien (2001) (with one child concept). For the purpose of representing this alignment, *Phyllotrox* sec. *auctorum* is an outgroup [76] with minimal relevance to merging the ingroup concepts. In order to limit such ingroup/outgroup interference, the input may be modified by (1) asserting 2006.PHY == 2001.PHY and (2) adding the letters "nc" – non-coverage – to the line "(2001.PHY 2001.PHYsubcin nc)" in the Franz & O'Brien (2001) input tree specification. Non-coverage means that the extension of parent concept 2001.PHY is only facultatively defined by its children. Using this annotation in combination with 2006.PHY == 2001.PHY has the intended effect of aligning the outgroup genus-level concepts congruently (figures 11 and 12).

Third, and most critically, the initial set of 17 input articulations in [1] includes 10 (of a total of 12) disjoint articulations among members of the *P. evelynae-P. sulcatae* clade sec. Franz & O'Brien (2001) on one side (five higher-level, five species-level) and the *P. carludovicae-P. evelynae* clade sec. Franz (2006) on the other side. This is so because concepts 2006.Pcar_Peve and 2001.Peve_Psul share co-extensional, synapomorphic *properties*. The properties ground the *intensional* articulations of congruence, or inclusion, respectively (e.g., 2006.Pcar_Peve == 2001.Peve_Psul [INT] and 2006.Pcar_Peve > 2001.Pvar [INT]). The ostensive reading, in turn, results in articulations of overlap or exclusion (e.g., 2006.Pcar_Peve >< 2001.Pbiv_Psul [OST] and 2006.Pcar_Peve | 2001.Pvar [OST]). By default the reasoner resolves each of the 12 disjoint, higher-level articulations in the ostensive sense (figure 11). The input of figure 11 yields 24 immediately deducible and 102 inferred articulations in the output MIR (Supporting Information S2). This input is overspecified; providing just eight species-level articulations for the ingroup concepts generates an identical merge (Table 1).

The containment with overlap graph (figure 11A) for the repaired, ostensive set of input articulations shows 15 overlapping articulations: six involving 2006.PHYLLO, five involving 2006.PER, and four involving 2006.Pcar_Peve. Each of these is an above species-level concept of the undersampled phylogeny sec. Franz (2006). The complexity of this and of the corresponding merge concept graph (figure 11B) is evident. The latter entails eight newly created Euler regions for which there are no congruent concepts in either input phylogeny.

In order to represent the intensional set of input articulations assessed in [1], four kinds of input adjustment are needed, as follows (compare figures 11 and 12). (1) An implied child 2006.Pcar_Peve_IC is introduced in the Franz (2006) phylogeny. (2) The 12 disjoint articulations of [1] are 'switched' from their ostensive interpretation to the intensional one; e.g. 2006.PER == 14 [INT], 2006.Pcar_Peve == 2001.Peve_Psul, or 2006.Pcar_Peve > 2001.Peve_Pvar [INT]. (3) The species-level concepts 2001.Pvariab, 2001.Pbivent, 2001.Psplend, 2001.Ppubico, and 2001.Psulcat must each be properly included in the implied child 2006.Pcar_Peve_IC (instead of being disjoint from 2006.Pcar_Peve as in figure 11). (4) The top-level articulation requires change from 2006.PHYLLO >< 2001.DER to 2006.PHYLLO == 2001.DER, given that 2006.PHY == 2001.PHY and 2006.PER == 2001.PER at the genus level. Based on these 17 modified input articulations, the reasoner produces a single merge using 29 immediately deducible and 115 inferred output MIR (Table 1).

---

[5] Even then this articulation is problematic; see results related to figure 12.

The introduction of the implied child 2006.Pcar_Peve_IC of *Perelleschus* sec. Franz (2006) and subsumption of five species-level concepts unique to Franz & O'Brien (2001) have the effect of this implied concept acting as an 'umbrella', or an asserted synapomorphy, under which the more densely sampled and finely resolved concepts of Franz & O'Brien (2001) can be represented. As a result, the containment with overlap analysis (figure 12A) has only four overlapping articulations in comparison to 15 such articulations under the ostensive reading of figure 11. The top-level tribal, generic, and informal clade concepts are all inferred as congruent, as opposed to each showing 4-6 overlapping articulations. The implied child 2006.Pcar_Peve_IC acts as the *differential* needed attain congruence among the 2001 clade-level concepts 2001.Peve_Pvar and 2001.Pbiv_Psul (and children of the latter) with the undersampled *P. carludovicae-P. evelynae* clade sec. Franz (2006). The representation asserts that 2006.Pcar_Peve == 2001.Peve_Psul, in the sense of having shared, co-extensional and synapomorphic properties, and in spite of differential sampling densities at lower taxonomic levels. The merge concept analysis (figure 12B) introduces four additional Euler regions nested under concepts 2001.Pbiv_Psul and 2006.Pcar_Peve_IC that resolve the differential more finely.

The intensional input articulations are overspecified; 10 primarily lower-level articulations are sufficient to produce the single, consistent alignment (Table 1).

**Alignment 5 – Franz (2006) and Franz & Cardona-Duque (2013)**

The alignment of the Franz (2006) and Franz & Cardona-Duque (2013) perspectives (figures 13 and 14) bears many similarities with alignment 4, and is therefore treated with brevity. The unaltered input of [1] includes several problematic constraints requiring repair. First, the non-coverage convention (nc) is needed to account for congruence of the non-focal outgroup concepts *Phyllotrox* sec. *auctorum*. However the input remains inconsistent after taking this action, and the reasoner indicates three repair options of which a change to 2013.PHYLLO == *or* > 2006.PHYLLO is suited to obtain the ostensive reading. Second, 2013.Peve_Pspi > 2006.Pcar_Peve must be modified to 2013.Peve_Pspi == *or* > 2006.Pcar_Peve to correctly represent the intensional reading – an omission in [1].

Following these changes, 18 disjoint articulations remain and are resolvable to yield unique merge taxonomies either under the ostensive (figures 13A and 13B) or intensional (figure 14A and 14B) reading, using representation solutions similar to those of alignment 4 (see also Table 1).

**Alignment 6 – Franz & O'Brien (2001) and Franz & Cardona-Duque (2013)**

The final alignment concerns the two broadly similar treatments of Franz & O'Brien (2001) and Franz & Cardona-Duque (2013). Both are phylogenetic revisions achieving comparable levels of clade resolution and terminal sampling. The most critical difference in the later treatment is the integration of two new species-level concepts: *P. salpinflexus* sec. Franz & Cardona-Duque (2013) and *P. spinothylax* sec. Franz & Cardona-Duque (2013). Jointly these entities constitute a monophyletic clade that is placed as sister to the *P. pubicoxae-P. sulcatae* clade sec. Franz & O'Brien (2001). In spite of the addition of lower-level members, several higher-level concepts pairs – i.e., concepts 2013.PER/2001.PER, 2013.Peve_Pspi/2001.Peve_Psul, 2013.Peve_Pvar/2001.Peve_Pvar, 2013.Pbiv_Pspi/2001.Pbiv_Psul, and 2013.Pcar_Pspi/2001.Pcar_Psul – remain intensionally congruent. Their properties, as proposed in the earlier (2001) revision, are reconfirmed by the authors of the later (2013) revision.

The unaltered input of [1] is inconsistent, due mainly to the original interpretation of articulations of two non-focal concept pairs. First, although each phylogeny lists only one congruent species-level concept (2013.PHYsubcin == 2001.PHYsubcin) as the child of *Phyllotrox* sec. *auctorum,* the articulation of the respective parent concepts is asserted as overlapping (2013.PHY >< 2001.PHY). This assertion is only appropriate under a more comprehensive representation of each parent concept [77]. Because these outgroup concepts are non-focal, we modify the input articulation to 2013.PHY == 2001.PHY. Second, the top-level input articulation asserts that the subtribal concept Phyllotrogina sec. Franz & Cardona-Duque (2013) overlaps with the tribal concept Derelomini sec. Franz & O'Brien (2001); i.e. 2013.PHYLLO >< 2001.DER. Again, this articulation could be validated through a more comprehensive representation of children under each parent concept; however for the present purpose this is unnecessary. Instead, we adjust the input to 2013.PHYLLO > 2001.DER for the ostensive reading (figure 15), and to 2013.PHYLLO == 2001.DER for the intensional reading (figure 16).

When given the original input of [1], the reasoner identifies inconsistency and suggests only one option for repair, i.e. the removal of the species-level articulation 2013.PHYsubcin == 2001.PHYsubcin. The repair yields a single possible world in which members of *Phyllotrox* sec. *auctorum* overlap variously with the higher-level concepts of the ingroups and with the newly added species-level concepts 2013.Psalpin and 2013.Pspinot. Even though this repair option involves fewer articulations than the 'manual' one described above, it is not appropriate. For further examination of this unintended repair see our Discussion.

Following modulation to assert the ostensive reading, the 20 input articulations produce a single merge based on 67 immediately deducible and 273 inferred output MIR (figure 15). Only 11 articulations are sufficient to generate this outcome. The containment with overlap graph (figure 15A) reflects not only the high degree of species-level concept congruence but also ten overlapping articulations at higher levels: four involving 2013.Pcar_Pspi, three involving 2013.Pbiv_Pspi, two involving 2013.Peve_Pspi, and one involving the genus-level concept 2013.PER. This succession is illuminating – in each case the later (2013) concept overlaps with an earlier (2001) concept 'ranked' (often informally) at the next inclusive level. The overlap originates at the lowest level with the intersection of 2013.Pcar_Pspi >< 2001.Pbiv_Psul, where the differential generated through the addition of new taxonomic elements by Franz & Cardona-Duque (2013) 'cascades up' the internal nodes of each phylogeny. Both 2013.Pcar_Psi and 2001.Pbiv_Psul share the equivalent of the *P. pubicoxae-P. sulcatae* clade sec. Franz & O'Brien (2001). This shared equivalent continues to increase with each higher level, terminating with 2013.PER and 2001.DER. The merge concept analysis (figure 15B) recognizes seven more finely resolved Euler regions created by overlapping articulations of the respective clade-level concept chains of the input phylogenies.

The alternative reading 2013.PHYLLO == 2001.DER requires adjusting seven disjoint articulations in [1] to the intensional setting (e.g., 2013.PER == 2001.PER, 2013.Psal_Pspi < 2001.Ppub_Psul). In addition, an implied child 2001.Ppub_Psul_**IC** must be introduced in the phylogeny of Franz & O'Brien (2001). As in previous alignments, the new species-level concepts sec. Franz & Cardona-Duque (2013) are asserted to be properly included within this implied child concept. The resulting 20 input articulations create a single merge based on 94 immediately deduced and 266 inferred articulations (figure 16). Twelve input articulations are sufficient to obtain the merge. The containment with overlap and merge concept analyses show identical merge taxonomies, where the congruent concept pair 2013.Psal_Pspi == 2001.Ppub_Psul_**IC** subsumes concepts 2013.Psalpin and 2013.Pspinot are unique to Franz & Cardona-Duque (2013). The remainder of the merge is entirely congruent, reflecting the reconfirmation of earlier (2001) property-centric clade concepts in the subsequent (2013) revision.

**Discussion**

In light of the detailed assessment of each *Perelleschus* alignment in the Results section, we direct the Discussion towards more general themes, as follows: (1) general implications of the concept alignment approach; (2) scalability constraints and prospects; and (3) future developments in reasoning, visualization, and integration.

Concepts, Articulations, and Reasoning – General Implications

**Representational complexity.** The inferred alignments for the *Perelleschus* use case are the first published demonstrations of reasoning over multiple taxonomies based on RCC-5 articulations and taking into account heterogeneous taxonomic constraints [6]. The *scale* of each alignment is small, ranging from 6-53 input concepts per pairwise alignment (Table 1). Nevertheless, the semantic *complexity* of this use case is considerable. In particular, the alignments succeed in logically representing and integrating numerous features of biological taxonomies where other approaches have so far failed [7,26,33,37,78–79]. Among these new achievements are: (1) compatibility with contemporary Linnaean nomenclature; (2) integration of many-to-many name/circumscription relationships across multiple input taxonomies; (3) reconciliation of traditional, ranked taxonomies with fully bifurcated and informally named phylogenies; (4) representation of monotypic concept lineages with multiple ranks yet congruent taxonomic extensions; (5) accounting for insufficiently specified higher-level entities, (6) undersampled outgroup entities, and (7) differentially sampled ingroup entities; (8) resolution of taxonomically overlapping entities and merge concepts; (9) differentiation of ostensive versus intensional, or hybrid readings of concept articulations; and (10) representation of topologically localized resolution versus ambiguity in alignments. In meeting these representation challenges, our approach surpasses preceding solutions to the challenge of multi-taxonomy alignment [12,33,40,49,79,80].

The use of Answer Set Programming is also novel in this context [81]. The combination of using high-performing reasoners [56] and polynomial encoding of the input [82] allows inferences of consistent alignments in efficient time. Answer Set Programming has the ability to solve non-monotonic reasoning problems and can directly represent disjoint RCC-5 articulations

[58–60,83]. Thus it is likely more suited for taxonomy alignment than the Description Logic (DL) approach which prevails in the Open Biomedical Ontologies domain [7,84–86].[6]

**Information gain.** The inferred alignments are consistent and thoroughly specified 'maps' according to which an input taxonomy may be integrated with its counterpart. Each of the alignment qualities – consistency, exhaustiveness, and expressiveness – is the outcome of the toolkit-enabled expert/reasoner interaction. This interaction is necessary because users are not guaranteed to produce well-specified alignments. As demonstrated above, the sets of input articulations provided by [1] have several shortcomings in this regard; including inconsistency, overspecification, or an inability to immediately represent intensional readings. Using the toolkit workflow can lead to the identification and resolution of these problems.

The reasoner-facilitated diagnoses, repairs, and modulations of the input also provide new insights into criteria of *sufficiency* for obtaining well-specified input alignments (Table 1). Accordingly, 39 or 41 non-disjoint articulations are sufficient to generate the respective ostensive or intensional alignments. When applied (with some redundancy) to the 13 analyses (figures 4–16), the set of 89 sufficient articulations logically entails 348 immediately deducible and 1081 inferred articulations, for a total of 1429 output MIR (Table 1). The input/output ratio amounts to a 16x gain in information, with the greatest increases obtained in the larger alignments. Thus, in addition to achieving logical consistency, the reasoning process *amplifies* the user's initial alignment effort. Even in small use cases such as the present one, the toolkit infers articulations an order of magnitude greater than the input.

**Knowledge integration.** The merge visualizations convey an immediate sense of taxonomic provenance and in-/stability. Generally speaking, if there is no difference between two taxonomies then we observe an isomorphic merge tree constituted only by grey squares ($T_1$ & $T_2$; under current visualization conventions). This is in fact the case for alignment 2 (figure 9) where only nomenclatural differences exist [1]. On the other hand, instances of non-congruent concepts are expressed through accumulations of yellow octagons ($T_1$) and green rectangles ($T_2$). Frequently (though not always) the *causes* for non-congruence are differential levels of taxonomic resolution, and are apparent in the visualizations as sets of low-level entities in one taxonomy that aggregate up to a single concept in the other taxonomy.

The *location* of congruent and unique concepts in the merge is also informative. One of the particularities of the *Perelleschus* use case is the absence of incongruent species-level concepts [1]: all instances of incongruence occur at higher levels. This outcome is readily derived from the merge visualizations (e.g., figure 15A). In addition, the extent and location of overlapping articulations indicate where regions of substantive disagreement or differential sampling exist [66,88–89].

Another advantage of the merge representations is their accessibility to computational agents and knowledge integration services [36,90]. The output MIR specify how to integrate taxonomic concepts and concept-associated information; they are *transmission graphs* that can propagate such information under logically consistent conditions [5,40,45–47]. Congruence (==) allows reciprocal information exchange. Inclusion (> or <) allows unidirectional data flow from the less to the more inclusive entity. Exclusion (|) prohibits such transmission. Overlap (><) is the most challenging articulation to integrate data over because it generates newly circumscribed Euler regions (compare figures 2E and 2F). In some instances the merge concept graph can resolve overlapping relations into finer entities, thereby reducing the challenge of data transmission to one of proper inclusion. Further research to integrate biodiversity data based on machine-interpretable merge taxonomies is clearly warranted.

---

[6] Issues of computational complexity and performance under different logic representations are further discussed in [43,51,82,87].

**Intensional and ostensive alignments.** Our use case is the first to differentiate between intensional and ostensive readings of taxonomic alignments [1,4,6–7,19]. However, we stress that the reasoner does not implement intensional readings directly. Such direct implementation would require representing the character information in the reasoning process. This is not possible with RCC-5 articulations, which act as proxies to such information, but can be achieved through other representation solutions [7,91–94].

By default the Euler/X toolkit reasoner provides ostensive alignments. In the 'eyes' of the logic reasoner, adding new species-level concepts to *Perelleschus* sec. Franz & Cardona-Duque (2013) alters and expands the circumscription *of cellular life*. Yet even though the option to produce bottom-to-top, ostension-based alignments is essential for comparing taxonomic content, this is not the only way to conceive of taxonomic equivalence [4, 6–7,26,74]. The notion that each additional species-level concept necessarily redefines the circumscription of all superseding parent concepts is counterintuitive to the way in which humans commonly conceive of taxonomic circumscriptions. We need reasoning approaches capable of representing both lower-level change and higher-level stability.

Alignment 6 illustrates this challenge. Under the ostensive reading (figure 15), the addition of the *P. salpinflexus-P. spinothylax* clade sec. Franz & Cardona-Duque (2013) expands the circumscription of every corresponding clade-level concept of Franz & O'Brien (2001). This outcome is valid but fails to reflect the re-/confirmed synapomorphic traits for these concepts in the 2001/2013 phylogenies. Introducing the implied child 2001.Ppub_Psul_**IC** and reconfiguring input articulations for the intensional reading (figure 16) does not alter the reasoning approach *per se*. Yet if effect these changes allow the implied concept to act *as if* representing the synapomorphic traits of the *P. pubicoxae-P. sulcatae* clade sec. Franz & O'Brien (2001). We thereby obtain an alignment that reflects shared, property-grounded intensionality along the internal nodes of each phylogeny. In summary, ostensive and intensional readings of alignments are both feasible and desirable because they reflect the hybrid nature of how taxonomic concepts contribute to human understanding of taxonomic content [73–75].

Scalability Constraints and Prospects

The challenge of resolving name/concept provenance in systematics has motivated a wide range of proposed solutions [12–15,33,78–80,95–97]. Our RCC-5/reasoning-based approach has strengths and limitations in this context. One theme of great relevance to wider adoption is scalability. We discuss several aspects of this theme, including human-, reasoner-, and resolution-specific scalability constraints and also prospects for overcoming these.

**Human constraints.** The beginning of the toolkit workflow depends on an expert's provision of input articulations. Unless such articulations are provided, no reasoning outcomes are attainable. Hence a wider implementation will likely depend on the adoption of new taxonomic annotation conventions [1] and web-based tools that facilitate the creation and reasoner-driven refinement of input articulations. At present no such tools are available; however the theoretical underpinnings for their production are well established [1–3,12,37–39,41,98–99].

The taxonomy alignment approach is about establishing mappings between different scientific theories of how certain names 'reach out' to perceived natural entities [73–74,100]. The theories are proposed and applied by human speakers, and accordingly the articulations reflect human-to-human linkages among the theories (that also happen to be directly interpretable by computational logic). Generating thousands to millions of new concept articulations without prior expert input is neither an option nor an objective of this approach. Instead, concept articulations are most fruitfully asserted in association with lower volume, high-quality systematic treatments, including legacy publications annotated by third-party experts. Use case analyses such as the present one may serve as blueprints for building better tools and thereby promote acceptance.

**Reasoning constraints.** Scalability trade-offs are also manifest in the reasoning process itself [45,51,54]. For the present, small-case study, all analyses were performed with a single 2.0 GHz processor and completed within less than one minute. Combining an innovative reasoning approach and polynomial encoding of the input constraints reduces the computational complexity of identifying all possible worlds [82]. Additional heuristic and parallelization strategies are nevertheless needed to handle use cases with many hundreds to thousands of input concepts.

Diagnosing inconsistencies becomes difficult when more than 3-5 'erroneous' articulations are simultaneously supplied to the reasoner [53–54]. One might alleviate this by allowing users to supply articulations sequentially for matching pairs of subtree regions. After adding a small number of articulations, the reasoner could perform interim consistency checks and repairs, leading to the incremental assembly of the complete alignment. This stepwise approach reduces the challenge of having to resolve multiple inconsistent articulations at once. Another option could involve assigning differential levels of certainty to input articulations and prioritize repair actions accordingly. Further formalization and application of minimal sufficiency criteria for achieving well-specified alignments would also lead to computational savings.

**Resolution constraints.** The *Perelleschus* use case has the advantage of having well defined outer boundaries and ingroup concepts [7]. These properties, in addition to the senior author's direct access to three of the six taxonomies, are conducive to obtaining a *single* possible world for each alignment and reading. Such a high degree of resolution will not always be attainable, however, which amounts to another constraint for this approach.

Explorations of underspecified input articulations for alignment 1 (figures 7–8) illustrate the effects of introducing ambiguity into the reasoning process. Larger use cases can generate vast numbers of possible worlds if the input is poorly specified. The toolkit includes aggregate and cluster graph options to visualize shared and unique properties among multiple alignments and resolve ambiguities where possible. However, discriminating among hundreds or thousands of logically equivalent merges remains challenging [101].

The degree of resolution of an alignment depends in part on the expressiveness of each input taxonomy, as exemplified in alignment 1. The vaguely circumscribed concept *Elleschus* sec. Günther (1936) permits three more or less plausible alignments (figures 4–6) to the succeeding concept *Perelleschus* sec. Voss (1954). Experts are challenged in such cases to represent ambiguity transparently while also producing expressive alignments. We suggest that many use cases will retain a measure of ambiguity in the resulting merge.

**Scalability prospects.** The herein illustrated approach amounts to one of the most powerful representations of taxonomic provenance developed to date. It overcomes systemic limitations of taxonomic names in identifying stability and change in taxonomic content while *retaining* all desirable features of contemporary nomenclatural practice. Concept articulations can represent member- or property-based equivalences of the respective concepts, without limiting the notion of identity – either in name or in circumscription – to type comparisons [102] or taxonomically insufficient criteria. In this sense reasoning over taxonomic concepts is also a viable complement to phylogenetic nomenclature [19,95,103].

Concept taxonomy has no ontological motivations to promote one or the other 'school' of naming in systematics. It is merely an epistemic approach to improve provenance tracking in cases were name/circumscription relationships change as an outcome of scientific advancement. Under the concept approach, 'valid' or 'reliable' or 'stable' names and circumscriptions are neither theoretically nor practically enforceable. We can neither conclusively predict nor restrict the taxonomic content of "*Perelleschus*" in the near or distant future. We *can* however align past and present usages of this name. Similarly, the purpose of aligning the concepts *Elleschus* sec. Günther (1936) and *Perelleschus* sec. Voss (1954) (figures 4–6) is not to show that one perspective is presently more valid than its counterpart. Instead, validity and reliability under the concept approach are only discernible facultatively and *a posteriori;* i.e., in cases where increasingly well supported systematic analyses produce chains of congruent concept articulations over extended time periods. Of course we do not mean to say that reliable and stable names are not desirable. We recognize, however, that the legacy of systematics is distinctly uneven in these regards. The focus of concept taxonomy is to build sound provenance chains amenable to computational representation and reasoning; *irrespective* of whether the nomenclatural and taxonomic history of a perceived lineage or organisms was 'perfect' since the times of Linnaeus or continues to experience major alterations.

In summary, we have shown that logically consistent alignments of heterogeneous taxonomies are feasible and improve upon alternative representations of taxonomic provenance. Further optimizations in computational and workflow performance are likely to overcome present scalability bottlenecks. The approach should merit wider application to advance biodiversity data representation in disciplines that benefit from these improvements [2–4,10–11,78–80,99,104].

## Future Developments – Reasoning, Visualization, and Integration

Future research to align taxonomies under the RCC-5/reasoning approach should focus on three main directions: (1) enhanced reasoning and visualization performance; (2) larger scale use case implementations; and (3) adoption of interactive web-based platforms that facilitate the taxonomy alignment workflow. Here we only discuss the first of these directions.

At present the alignments are restricted to two input taxonomies ($T_1,T_2$), though it is possible in principle to first merge $T_1$ and $T_2$ into a single taxonomy ($T_{1\bullet 2}$) and then articulate the merge with an additional taxonomy ($T_3$). For instance, one could merge the congruent taxonomies of Voss (1954) and Wibmer & O'Brien (1986) (figure 9) with the intensional merge of Franz & O'Brien (2001) and Franz & Cardona-Duque (2013) (figure 16). Such a four-taxonomy alignment outcome will bear similarities with alignment 3 (figure 10). Future research should develop solutions for aligning three or more taxonomies, either sequentially or simultaneously [40].

The merge visualizations are also in need of optimization [79,101,105]; in particular with regards to the interaction of containment and merge concept views (e.g., figures 2E and 2F). Although the former are perhaps easier to navigate by humans, the latter resolve concept overlap more accurately (e.g., figures 10–15). An improved visualization application would allow users to explore overlap dynamically, by selecting overlapping articulations and obtaining a 'zoom-in' view with merge concept resolution. The labeling conventions and hierarchical interactions of merge concepts also require refinement.

Lastly, we reiterate that concept articulations and alignments should be grounded as transparently as possible in high-quality information about taxonomic names [14], concepts [1,9,23], occurrences [99], and their phenotypic and genotypic properties [1,7,92–94]. Integration of these interdependent information sources is essential to linking the asserted articulations to systematic data that remain amenable to disagreement and reexamination. The articulations themselves are also linked to authors and times of creation. They are not 'objective', although they can be inter-subjectively validated, or challenged. A central objective of the concept approach is to increase the transparency of provenance among high-quality systematic treatments and expert perspectives.

## Supporting Information

**File S1    Set of toolkit input files for 14 alignments shown in figures 2 and 4-16.** Each input file, saved in .txt format, contains annotations and instructions for run commands to yield the alignments and visualizations shown in the 14 corresponding figures.

**File S2    Set of toolkit output MIR (maximally informative relations) files for the 14 alignments shown in figures 2 and 4-16.** Each output file, saved in .csv format, is sorted according to the deduced/inferred output MIR. See also Supporting Information S1 and main text.

The entire *Perelleschus* use case has been deposited as an experiment for access and reproduction at http://recomputation.org/

## Acknowledgments

The authors thank Charles O'Brien and Juliana Cardona-Duque for their taxonomic contributions to the *Perelleschus* use case. Supporting for the authors' research through the National Science Foundation is kindly acknowledged (awards DEB–1155984 and DBI–1342595 to NMF; and IIS–118088 and DBI–1147273 to BL).

## Author Contributions

Conceived and designed the experiments: NMF. Performed the experiments: NMF BL MC PK SB SY. Analyzed the data: NMF BL MC PK SY. Contributed reagents/materials/analysis tools: NMF BL MC PK SB SY. Wrote the paper: NMF BL MC.

## References

1.  Franz NM, Cardona-Duque J (2013) Description of two new species and phylogenetic reassessment of *Perelleschus* Wibmer & O'Brien, 1986 (Coleoptera: Curculionidae), with a complete taxonomic concept history of *Perelleschus* sec. Franz & Cardona-Duque, 2013. Syst. Biodivers. 11: 209–236.
2.  Berendsohn WG (1995) The concept of "potential taxa" in databases. Taxon 44: 207–212.
3.  Kennedy J, Kukla R, Paterson T (2005) Scientific names are ambiguous as identifiers for biological taxa: their context and definition are required for accurate data integration. In:

Ludäscher B, & Raschid L, eds. Data Integration in the Life Sciences: Proceedings of the Second International Workshop, San Diego, CA, USA, July 20–22. DILS 2005, LNBI 3615. pp. 80–95.
4. Franz NM, Peet RK, Weakley AS (2008) On the use of taxonomic concepts in support of biodiversity research and taxonomy. In: Wheeler QD, ed. The New Taxonomy, Systematics Association Special Volume Series 74. Boca Raton: Taylor & Francis. pp. 63–86.
5. Geoffroy M, Berendsohn WG (2003) Transmission of taxon-related factual information. Schrift. Vegetationsk. 39: 83–86.
6. Franz NM, Peet RK (2009) Towards a language for mapping relationships among taxonomic concepts. Syst. Biodivers. 7: 5–20.
7. Franz NM, Thau D (2010) Biological taxonomy and ontology development: scope and limitations. Biodiversity Informatics 7: 45–66.
8. Franz NM, Chen M, Yu S, Bowers S, Ludäscher B (2014) Names are not good enough: reasoning over taxonomic change in the *Andropogon* complex. Semantic Web – Interoperability, Usability, Applicability – Special Issue on Semantics for Biodiversity. (In Review)
9. Koperski M, Sauer M, Braun W, Gradstein SR (2000) Referenzliste der Moose Deutschlands. Schrift. Vegetationsk. 34: 1–519.
10. Vane-Wright RI (2003) Indifferent philosophy versus almighty authority: on consistency, consensus and unitary taxonomy. Syst. Biodivers. 1: 3–11.
11. Franz NM (2005) On the lack of good scientific reasons for the growing phylogeny/classification gap. Cladistics 21: 495–500.
12. Tuominen J, Laurenne N, Hyvönen E (2011) Biological names and taxonomies on the Semantic Web – managing the change in scientific conception. In: Antoniou G, Grobelnik M, Simperl E, Parsia B, Plexousakis D, Leenheer P, Pan J, eds. The Semantic Web: Research and Applications. Lect. Notes Comput. Sc. 6644: 255–269.
13. Schuh RT (2003) The Linnaean system and its 250-year persistence. Bot. Rev. 69: 59–78.
14. Patterson DJ, Cooper J, Kirk PM, Pyle RL, Remsen DP (2010) Names are key to the big new biology. Trends Ecol. Evol. 25: 686–691.
15. Page RDM (2013) BioNames: linking taxonomy, texts, and trees. PeerJ 1: e190.
16. Cheney J, Chiticariu L, Tan W-C (2007) Provenance in databases: why, how, and where. Foundations and Trends in Databases 1: 379–474.
17. Moreau L (2010) The foundations for provenance on the Web. Foundations and Trends in Web Science 2: 99–241.
18. de Paula R, Holanda M, Gomes LSA, Lifschitz S, Walter MAMT (2013) Provenance in bioinformatics workflows. BMC Bioinformatics 14(Suppl. 11): S6.
19. Franz NM (2009) Letter to Linnaeus. In: Knapp S, Wheeler QD, eds. Letters to Linnaeus. London: Linnean Society of London. pp. 63–74.
20. Fabricius JC (1792) Entomologia systematica emendata et aucta. Secundum classes, ordines, genera, species adjectis synonimis, locis, observationibus, descriptionibus. Vol. 1. Hafniae: Proft. XX + 538 pp.
21. Bouchard P, Bousquet Y, Davies AE, Alonso-Zarazaga MA, Lawrence JF, Lyal CHC, Newton AF, Reid CAM, Schmitt M, Ślipiński SA, Smith ABT (2011) Family-group names in Coleoptera (Insecta). ZooKeys 88: 1–972.
22. Schuh RT, Brower AVZ (2009) Biological Systematics: Principles and Applications. Second Edition. Ithaca: Cornell University Press. 311 pp.

23. Weakley AS (2012) Flora of the Southern and Mid-Atlantic States. Working Draft of 30 November 2012. 1225 pp. Available: http://www.herbarium.unc.edu/FloraArchives/WeakleyFlora_2012-Nov.pdf Accessed 2014 March 30.
24. Midford PE, Dececchi TA, Balhoff JP, Dahdul WM, Ibrahim N, Lapp H, Lundberg JG, Mabee PM, Sereno PC, Westerfield M, Vision TJ, Blackburn DC (2013) The vertebrate taxonomy ontology: a framework for reasoning across model organism and species phenotypes. J. Biomed. Sem. 4: 34.
25. Bizer C, Heath T, Berners-Lee T (2009) Linked Data – the story so far. Int. J. Semant. Web Inf. Syst. 5: 1–22.
26. Schulz S, Stenzhorn H, Boeber M (2008) The ontology of biological taxa. Bioinformatics 24: i313–i321.
27. Page RDM (2006) Taxonomic names, metadata, and the Semantic Web. Biodiversity Informatics 3: 1–15.
28. Page RDM (2008) Biodiversity informatics: the challenge of linking data and the role of shared identifiers. Brief. Bioinform. 9(5): 345–354.
29. ICZN – International Commission on Zoological Nomenclature (1999) International Code of Zoological Nomenclature. Fourth Edition. London: International Trust for Zoological Nomenclature. 306 pp.
30. McNeill J, Turland NJ, Barrie FR, Buck WR, Greuter W, Wiersema JH (2012) International Code of Nomenclature for Algae, Fungi, and Plants (Melbourne Code). Königstein: Koeltz Scientific Books. 208 pp.
31. ICSP – International Committee on Systematics of Prokaryotes. Available: http://icsp.org/ Accessed 2014 March 30.
32. McCarthy J (1989) Artificial intelligence, logic and formalizing common sense. In: Thomason RH, ed. Philosophical Knowledge and Artificial Intelligence. Springer: New York. pp. 161–190.
33. Chawuthai R, Takeda H, Wuwongse V, Jinbo U (2013) A logical model for taxonomic concepts for expanding knowledge using Linked Open Data. In: Larmande P, Arnaud E, Mougenot I, Jonquet C, Libourel T, Ruiz M, eds. S4BioDiv 2013, Semantics for Biodiversity – Proceedings of the First International Workshop on Semantics for Biodiversity, Montpellier, France, May 27, 2013. CEUR Workshop Proceedings, Vol-797. pp. 1–8.
34. Tschöpe O, Macklin JA, Morris RA, Suhrbier L, Berendsohn, WG (2013) Annotating biodiversity data via the Internet. Taxon 62: 1248–1258.
35. Gradstein SR, Sauer M, Koperski M, Braun W, Ludwig G (2001) TaxLink, a program for computer-assisted documentation of different circumscriptions of biological taxa. Taxon 50 1075–1084.
36. Euzenat J, Shvaiko P (2013) Ontology Matching. Second Edition. New York: Springer.
37. Beach JH, Pramanik S, Beaman JH (1993) Hierarchic taxonomic databases. In: Fortuner R, ed. Advances in Computer Methods for Systematic Biology: Artificial Intelligence, Databases,Computer Vision. Baltimore: John Hopkins University. pp. 241–256.
38. Berendsohn WG (1997) A taxonomic information model for botanical databases: the IOPI Model. Taxon 46: 283–309.

39. Koperski M, Sauer M (1999) Das Projekt "Referenzliste der Moose Deutschlands" – Dokumentation unterschiedlicher taxonomischer Auffassungen mit Hilfe des Datenbankprogrammes TAXLINK. Stuttgarter Beitr. Naturk., Serie A, 590: 1–10.
40. Geoffroy M, Güntsch A (2003) Assembling and navigating the potential taxon graph. Schrift. Vegetationsk. 39: 71–82.
41. Berendsohn WG, ed. (2003) MoReTax: handling factual information linked to taxonomic concepts in biology. Schrift. Vegetationsk.: 39: 1–113.
42. Randell DA, Cui Z, Cohn AG (1992) A spatial logic based on regions and connection. In: Nebel B, Swartout W, Rich C, eds. Proceedings of the Third International Conference on the Principles of Knowledge Representation and Reasoning. Los Altos: Morgan Kaufmann. pp. 165–176.
43. Thau D, Ludäscher B (2007) Reasoning about taxonomies in first-order logic. Ecol. Inform. 2: 195–209.
44. Thau D (2008) Reasoning about taxonomies and articulations. In: Ph.D. '08: Proceedings of the 2008 EDBT Ph.D. Workshop. New YorK: ACM. pp. 11–19.
45. Thau D, Bowers S, Ludäscher B (2008) Merging taxonomies under RCC-5 algebraic articulations. In: Conference on Information and Knowledge Management. Proceeding of the 2nd International Workshop on Ontologies and Information systems for the Semantic Web. New York: ACM. pp. 47–54.
46. Thau D, Bowers S, Ludäscher B (2009) Merging sets of taxonomically organized data using concept mappings under uncertainty. In: Proceedings of the 8th International Conference on Ontologies, Databases, and the Applications of Semantics (ODBASE 2009). OTM 2009. Lect. Notes Comput. Sci. 5871: 1103–1120.
47. Thau D, Bowers S, Ludäscher B (2010) Towards best-effort merge of taxonomically organized data. In: 2010 IEEE 26th International Conference on Data Engineering Workshops (ICDEW). pp. 151–154.
48. Ferreirós J (2001) The road to modern logic – an interpretation. Bull. Symb. Log. 7: 441–484.
49. Thau D, Bowers S, Ludäscher B (2009) CleanTax: a framework for reasoning about taxonomies. Proceedings of the AAAI Spring Symposium – Benchmarking of Qualitative Spatial and Temporal Reasoning Systems. pp. 49–50.
50. McCune W (2010) Prover9 and Mace4. Available: www.cs.unm.edu/~mccune/prover9 Accessed 2014 March 30.
51. Chen M, Yu S, Franz N, Bowers S, Ludäscher B (2014) Euler/X: a toolkit for logic-based taxonomy integration. arXiv:1402.1992 [cs.LO] Available: http://arxiv.org/abs/1402.1992 Accessed 2014 March 30.
52. Chen M (2014) Euler Project source code. Available: https://bitbucket.org/eulerx/euler-project Accessed 2014 March 30.
53. Chen M, Bowers S, Yu S, Franz N, Ludäscher B (2014) Using black-box and white-box provenance for explaining taxonomy alignments. In: TaPP'14 – 6th USENIX Workshop on the Theory and Practice of Provenance. (In Review)
54. Chen M, Yu S, Franz N, Bowers S, Ludäscher B (2014) A hybrid diagnosis approach combining black-box and white-box approaches. In: NMR 2014 – 15th International Workshop on Non-Monotonic Reasoning. (In Review)
55. DLVSYSTEM (2014) Available: http://www.dlvsystem.com/ Accessed 2014 March 30.

56. Gebser M, Kaufmann B, Kaminski R, Ostrowski M, Schaub T, Schneider MT (2011) Potassco: the Potsdam Answer Set Solving Collection. AI Commun. 24: 107–124.
57. Gansner ER, North SC (2000) An open graph visualization system and its applications to software engineering, Soft. Pract. Exper. 30: 1203–1233.
58. Lifschitz V (2008) Twelve definitions of a stable model. In: de la Banda MG, Pontelli E, eds. Logic Programming. Lect. Notes Comput. Sci. 5366: 37–51.
59. Gelfond M (2008) Answer sets. In: van Harmelen F, Lifschitz V, Porter B, eds. Handbook of Knowledge Representation. Amsterdam: Elsevier. pp. 285–316.
60. Brewka G, Eiter T, Truszczyński M (2011) Answer set programming at a glance. Commun. ACM 54: 92–103.
61. Wibmer GJ, O'Brien CW (1986) Annotated checklist of the weevils (Curculionidae *sensu* lato) of South America (Coleoptera: Curculionoidea). Mem. Amer. Entomol. Inst. 39: 1–563.
62. Franz NM (2006) Towards a phylogenetic system of derelomine flower weevils (Coleoptera: Curculionidae). Syst. Entomol. 31: 220–287.
63. Page RDM (2011) Dark taxa: GenBank in a post-taxonomic world. Available: http://iphylo.blogspot.com/2011/04/dark-taxa-genbank-in-post-taxonomic.html Accessed 2014 March 30.
64. Bryant D (2003) A classification of consensus methods for phylogenies. In: Janowitz M, Lapointe F-J, McMorris FR, Mirkin B, Roberts FS, eds. BioConsensus, DIMACS. AMS. pp. 163–184.
65. Cha S, Tappert CC (2009) A genetic algorithm for constructing compact binary decision trees," J. Patt. Recog. Res. 4: 1–13.
66. Sangster G (2014) The application of species criteria in avian taxonomy and its implications for the debate over species concepts. Biol. Rev. 89: 199–2014.
67. Howse J, Stapleton G, Flower J, Taylor J (2002) Corresponding regions in Euler diagrams. In: Diagrammatic Representation and Inference. Lect. Notes Comput. Sci. 2317: 76–90.
68. Günther K (1936) Notizen über Rüsselkäfer aus Costa Rica. Entomol. Rund. 53: 190–192.
69. Voss E (1954) Curculionidae (Col.). In: Titschack E, ed. Beiträge zur Fauna Perús, Band IV, Wissenschaftliche Bearbeitungen. Jena: VEB Gustav Fischer Verlag. pp. 193–376.
70. Franz NM, O'Brien CW (2001) Revision and phylogeny of *Perelleschus* (Coleoptera: Curculionidae), with notes on its association with *Carludovica* (Cyclanthaceae). Trans. Amer. Entomol. Soc. 127: 255–287.
71. Prendini L (2001) Species or supraspecific taxa as terminals in cladistic analysis? Groundplans versus exemplars revisited. Syst. Biol. 50: 290–300.
72. Lacordaire T (1863) *Histoire naturelle des insectes. Genera des coléoptères, ou exposé méthodique et critique de tous les genres proposés jusqu'ici dans cet ordre d'insectes, vol. 6.* Paris: Roret.
73. Devitt M, Sterelny K (1999) Language and Reality: an Introduction to the Philosophy of Language. Second Edition. Cambridge: MIT Press.
74. Brigandt, I. 2009. Natural kinds in evolution and systematics: metaphysical and epistemological considerations. Acta Biotheor. 57: 77–97.
75. Schulz S, Stenzhorn H, Boeber M (2008) The ontology of biological taxa. Bioinformatics 24: i313– i321.
76. Nixon KC, Carpenter JM (1993) On outgroups. Cladistics 9: 413–426.

77. Franz NM (2003) Systematics of *Cyclanthura,* a new genus of Derelomini (Coleoptera: Curculionidae). Insect Syst. Evol. 34: 153–198.
78. Thomson RC, Shaffer HB (2010) Sparse supermatrices for phylogenetic inference: taxonomy, alignment, rogue taxa, and the phylogeny of living turtles. Syst. Biol. 59: 42–58.
79. Smith SA, Brown JW, Hinchcliff CE (2013) Analyzing and synthesizing phylogenies using tree alignment graphs. PLoS Comp. Biol. 9(9): e1003223.
80. Boyle B, Hopkins, Lu Z, Raygoza Garay JA, Mozzherin D, Rees T, Matasci N, Narro ML, Piel WH, Mckay SJ, Lowry S, Freeland C, Peet RK, Enquist BJ (2013) The taxonomic name resolution service: an online tool for automated standardization of plant names. BMC Bioinform. 2013, 14:16.
81. Erdem E (2011) Applications of Answer Set Programming in phylogenetic systematics. In: Balduccini M, Son TC, eds. Logic Programming, Knowledge Representation, and Nonmonotonic Reasoning. Lect. Notes Comput. Sci. 6565: 415–431.
82. Chen M (2014) Query Optimization and Taxonomy Integration. Ph.D. Dissertation, University of California at Davis.
83. Baral C, Gelfond M, Rushton N (2004) Probabilistic reasoning with answer sets. In: Lifschitz V, Niemelä I, eds. Proceedings of the 7th International Conference of Logic Programming and Nonmonotonic Reasoning, Fort Lauderdale, FL. Lect. Notes Comput. Sci. 2923: 21–33.
84. Smith B, Ceusters W, Klagges B, Köhler J, Kumar A, Lomax J, Mungall C, Neuhaus F, Rector AL, Rosse C (2005) Relations in biomedical ontologies. Genome Biology 2005, 6:R46.
85. Horrocks I (2008) Ontologies and the Semantic Web. Comm. ACM 51: 58–67.
86. Lukasiewicz T (2010) A novel combination of Answer Set Programming with Description Logics for the Semantic Web. IEEE Trans. Knowl. Data Eng. 22: 1577–1592.
87. Thau DM (2010) Reasoning about Taxonomies. Ph.D. Dissertation, University of California at Davis.
88. Meiri S, and G.M. Mace (2007) New taxonomy and the origin of species. PLoS Biology 5(7): e194.
89. Padial JM, De La Riva I (2007) Taxonomic inflation and the stability of species lists: the perils of ostrich's behavior. Syst. Biol. 55: 859–867.
90. Van Harmelen F, Lifschitz V, Porter B, eds. (2008) The Handbook of Knowledge Representation. Elsevier: Oxford.
91. Dahdul WM, Lundberg JG, Midford PE, Balhoff JP, Lapp H, Vision TJ, Haendel MA, Westerfield M, Mabee (2010) The Teleost Anatomy Ontology: anatomical representation for the genomics age. Syst. Biol. 59: 369–383.
92. Mungall CJ, Gkoutos GV, Smith CL, Haendel MA, Lewis SE, Ashburner M (2010) Integrating phenotype ontologies across multiple species. Genome Biology 2010, 11:R2.
93. Cui H (2012) CharaParser for fine-grained semantic annotation of organism morphological descriptions. J. Am. Soc. Inf. Sci. Technol. 63: 738–754.
94. Deans AR, Yoder MJ, Balhoff JP (2012) Time to change how we describe biodiversity. Trends Ecol. Evol. 27: 78–84.
95. De Queiroz K, Gauthier J (1992) Phylogenetic taxonomy. Annu. Rev. Ecol. Syst. 23: 449–480.
96. Midford PE, Dececchi TA, Balhoff JP, Dahdul WM, Ibrahim N, Lapp H, Lundberg JG, Mabee PM, Sereno PC, Westerfield M, Vision TJ, Blackburn DC (2013). The vertebrate taxonomy ontology: a framework for reasoning across model organism and species phenotypes. J. Biomed. Sem. 4: 34.
97. Marakeby H, Badr E, Torkey H, Song Y, Leman S, Monteil CL, Heath LS, Vinatzer BA (2014) A system to automatically classify and name any individual genome-sequenced organism independently of current biological classification and nomenclature. PLoS ONE 9(2): e89142.
98. Kennedy J, Hyam R, Kukla R, Paterson T (2006) Standard data model representation for taxonomic information. OMICS 10: 220–230.
99. Baskauf SJ, Webb CO (2014) Darwin-SW: Darwin Core-based terms for expressing biodiversity data as RDF. Semantic Web – Interoperability, Usability, Applicability – Special Issue on Semantics for Biodiversity. (In Review)
100. Reimer M (2010) Reference. In: Zalta EN, ed. The Stanford Encyclopedia of Philosophy (Spring 2010 Edition). Available: http://plato.stanford.edu/archives/spr2010/entries/reference/ Accessed 2014 March 30.
101. Graham M, Kennedy J (2010) A survey of multiple tree visualisation. Inform. Vis. 9: 235–252.
102. Witteveen J (2014) Naming and contingency: the type method of biological taxonomy. Biol. Phil. (in press)
103. Bryant HN, Cantino PD (2002) A review of criticisms of phylogenetic nomenclature: is taxonomic freedom the fundamental issue? Biol. Rev. 77: 39–55.
104. Vines TH, Albert AYK, Andrew RL, Débarre F, Bock DG, Franklin MT, Gilbert KJ, Moore J-S, Renaut S, Rennison DJ (2014) The availability of research data declines rapidly with article age. Curr. Biol. 24: 94–97.
105. Paradis E, Claude J, Strimmer K (2004) APE: analyses of phylogenetics and evolution in R language. Bioinform. 20: 289–290.

**Figure Captions**

**Figure 1. Reasoning workflow schema.** To initiate the workflow, two input taxonomies ($T_1$,$T_2$) are supplied jointly with a set of concept articulations (A) and taxonomic constraints (C).
The workflow facilitates an iterative alignment process aimed at rendering the input logically consistent and sufficiently expressive. Negation of either criterion (red arrows) leads to input modulation through either diagnosis/removal of conflicting constraints (no possible worlds; right loop) or exploration of many possible worlds and provision of additional constraints (left loop). The well-specified alignment (green arrows) is output as a set of MIR (maximally informative relations; both immediately deducible and inferred [ded./inf.]), and visualized either as a containment (with overlap [><]) or merge concept graph (see figure 2).

**Figure 2. Abstract toolkit input and output example.** (A) Input taxonomy $T_1$, with nine concepts named (1.) A-I. (B) Input Taxonomy $T_2$, with eight concepts named (2.) A-I. Concept 2.CD is congruent with (1.C + 2.D). The respective child concepts 1.E/2.E and 1.G/2.G are non-congruently assigned to parent concepts 1.B/2.B and 1.F/2.F. (C) Representation of $T_1$ and $T_2$ and articulations in the toolkit input file (see also Supporting Information S1). (D) Toolkit input visualization, showing both hierarchical (intra-taxonomic; *is_a*) and lateral (inter-taxonomic; RCC-5) articulations. (E) Single, consistent alignment of the input shown as a containment with overlap graph (legend to the right of [D]). (F) Merge concept analysis of the input, resolving Euler regions that result from overlapping concepts. Annotation convention: 1.B\2.B = the region of the 1.B/2.B overlap which is unique to 1.B; 1.B*2.B = the region which each concept shares; and 2.B\1.B = the region which is unique to 2.B.

**Figure 3. Input taxonomies for the *Perelleschus* use case.** All 54 concepts are uniquely identified, and labeled with either traditional (ranked) or informal (clade) names. Concept label abbreviations are provided in square brackets and are used throughout this analysis.

**Figure 4. Alignment 1 – Günther (1936 – $T_1$) and Voss (1954 – $T_2$), ostensive reading [OST].** Input articulations are shown on the left; only the articulation 1954.PER > 1936.ELL is represented (as inferred) in the output MIR (maximally informative relations). * = sufficient input articulations – this annotation convention (*) is used in all subsequent figures where sufficiency is obtained with a subset of the input articulations. The containment with overlap and merge concept graphs are identical.

**Figure 5. Alignment 1 – Günther (1936 – $T_1$) and Voss (1954 – $T_2$), intensional/ostensive reading [INT/OST].** Modifications of the input alignment in comparison to that of figure 4 are shown in bold font. Other conventions as specified in figures 2-4. The top-level articulation 1954.PER >< 1936.ELL is provided, and an implied child 1936.ELL_IC is introduced, where 1954.PER | 1936.ELL_IC. (A) Containment with overlap graph. (B) Merge concept graph.

**Figure 6. Alignment 1 – Günther (1936 – $T_1$) and Voss (1954 – $T_2$), intensional reading [INT].** The top-level articulations is asserted as 1954.PER < 1936.ELL and an implied child 1936.ELL_IC is introduced. (A) Containment with overlap graph. (B) Merge concept graph.

**Figure 7. Alignment 1 – Günther (1936 – $T_1$) and Voss (1954 – $T_2$), intensional/ostensive reading [INT/OST]), underspecified input, level 1 (one articulation removed).** Input constraints as in figure 5, yet without 1954.PER | 1936.ELL_IC, resulting in five disjoint articulations involving concepts 1936.ELL and 1936.ELL_IC in the inferred output MIR. (A-H) Containment with overlap graphs of eight consistent alignments (possible worlds).

**Figure 8. Alignment 1 – Günther (1936 – $T_1$) and Voss (1954 – $T_2$), intensional/ostensive reading [INT/OST]), underspecified input, level 2 (two articulations removed).** Input configuration as in figure 6, yet without 1954.PER >< 1936.ELL, resulting in six disjoint articulations in the inferred output MIR. (A-I) Containment with overlap graphs of nine *additional* alignments beyond those shown in figure 7, resulting in a total of 17 possible worlds.

**Figure 9. Alignment 2 – Voss (1954 – $T_1$) and Wibmer & O'Brien (1986 – $T_2$), intensional/ostensive reading [INT/OST].** Conventions as in figure 4; the output MIR are provided in the Supporting Information S2.

**Figure 10. Alignment 3 – Wibmer & O'Brien (1986 – $T_1$) and Franz & O'Brien (2001 – $T_2$), intensional/ostensive reading [INT/OST].** The implied child 2001.PHY_IC of parent 2001.PHY is the only input modification in comparison to [1]. (A) Containment with overlap graph. (B) Merge concept graph.

**Figure 11. Alignment 4 – Franz & O'Brien (2001 – $T_1$) and Franz (2006 – $T_2$), ostensive reading [OST].** Modifications of the initial input given in [1] are shown in bold font. Non-coverage (nc) is asserted for parent concept 2006.PHY of child concept 2006.PHYsubcin. (A) Containment with overlap graph. (B) Merge concept graph.

**Figure 12. Alignment 4 – Franz & O'Brien (2001 – $T_1$) and Franz (2006 – $T_2$), intensional reading [INT].** See also figure 11. The implied child 2006.Pcar_Peve_IC is introduced and asserted to include five species-level concepts sec. Franz & O'Brien (2001). (A) Containment with overlap graph. (B) Merge concept graph.

**Figure 13. Alignment 5 – Franz (2006 – $T_1$) and Franz & Cardona-Duque (2013 – $T_2$), ostensive reading [OST].** Modifications of the initial input given in [1] are shown in bold font. Non-coverage (nc) is stipulated for parent concept 2013.PHY of child concept 2013.PHYsubcin. (A) Containment with overlap graph. (B) Merge concept graph.

**Figure 14. Alignment 5 – Franz (2006 – $T_1$) and Franz & Cardona-Duque (2013 – $T_2$), intensional reading [INT].** See also figure 13. The implied child 2006.Pcar_Peve_IC (2006:Pis introduced and asserted to include seven species-level concepts sec. Franz & Cardona-Duque (2013). (A) Containment with overlap graph. (B) Merge concept graph.

**Figure 15. Alignment 6 – Franz & O'Brien (2001 – $T_1$) and Franz & Cardona-Duque (2013 – $T_2$), ostensive reading [OST].** Modifications of the initial input given in [1] are shown in bold font. (A) Containment with overlap graph. (B) Merge concept graph.

**Figure 16. Alignment 6 – Franz (2006 – T$_1$) and Franz & Cardona-Duque (2013 – T$_2$), intensional reading [INT].** See also figure 15. The implied child Ppub_Psul_**IC** is introduced and asserted to include two species-level concepts sec. Franz & Cardona-Duque (2013). The containment with overlap and merge concept graphs are identical.

**Table 1.** Summary of input concepts and input/output articulations (MIR – maximally informative relations) for the *Perelleschus* use case, corresponding to the 13 readings for alignments 1-6 shown in figures 4-16. Numbers of *sufficient* input articulations (marked as * in the respective figures) are provided; with the initial numbers of [1] shown in parentheses.

| Alignment | T$_1$ – T$_2$ | Reading | Figure | Concepts | Articulations | MIR-Deduced | MIR-Inferred | MIR-Total |
|---|---|---|---|---|---|---|---|---|
| 1 | 1936 – 1954 | OST | 4 | 6 | 3 (4) | 4 | 4 | 8 |
| 1 | 1936 – 1954 | INT/OST | 5 | 7 | 5 | 7 | 5 | 12 |
| 1 | 1936 – 1954 | INT | 6 | 7 | 4 | 6 | 6 | 12 |
| 1 | 1936 – 1954 | INT/OST | 7 | 7 | 4 | 6 | 6 | 12 |
| 1 | 1936 – 1954 | INT/OST | 8 | 7 | 3 | 5 | 7 | 12 |
| 2 | 1954 – 1986 | INT/OST | 9 | 8 | 3 (4) | 12 | 4 | 16 |
| 3 | 1986 – 2001 | INT/OST | 10 | 29 | 3 (7) | 12 | 60 | 72 |
| 4 | 2001 – 2006 | OST | 11 | 33 | 8 (17) | 24 | 102 | 126 |
| 4 | 2001 – 2006 | INT | 12 | 34 | 10 | 29 | 115 | 144 |
| 5 | 2006 – 2013 | OST | 13 | 37 | 11 (23) | 20 | 127 | 147 |
| 5 | 2006 – 2013 | INT | 14 | 38 | 12 | 62 | 106 | 168 |
| 6 | 2001 – 2013 | OST | 15 | 52 | 11 (20) | 67 | 273 | 340 |
| 6 | 2001 – 2013 | INT | 16 | 53 | 12 | 94 | 266 | 360 |
| **Totals** | – | – | – | **318** | **89 (125)** | **348** | **1081** | **1429** |

# Figure 1

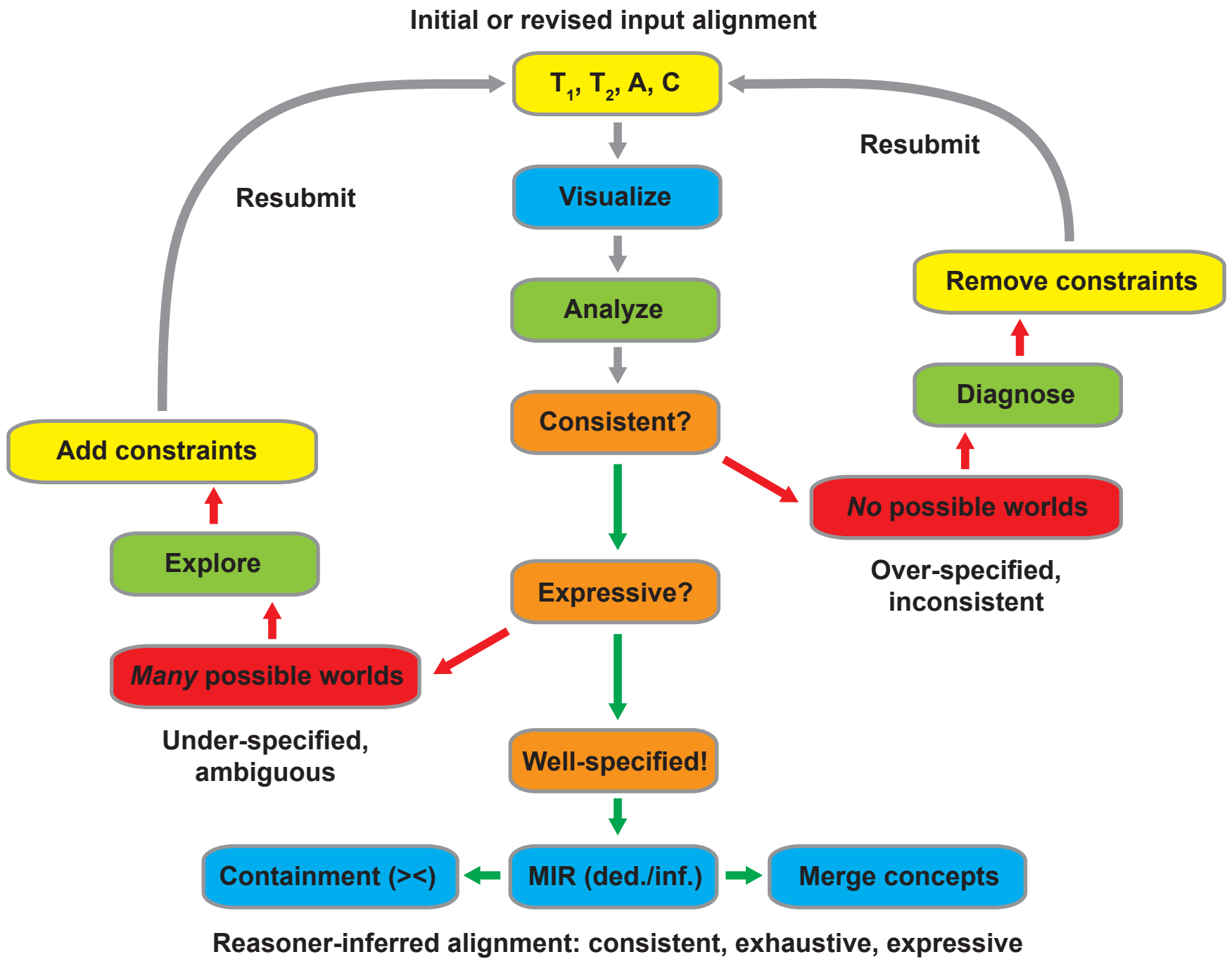

Figure 2

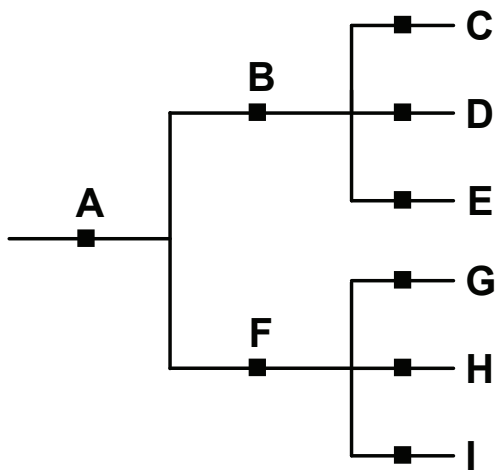

A  Taxonomy 1 (T₁)

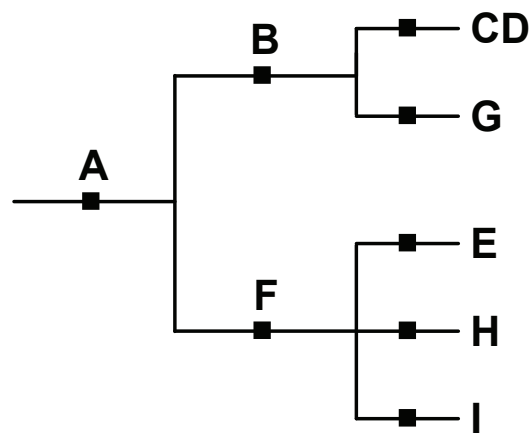

B  Taxonomy 2 (T₂)

C  Articulations (A)

**taxonomy 2**
(A B F)
(B CD G)
(F E H I)

**taxonomy 1**
(A B F)
(B C D E)
(F G H I)

**articulation figure2**
2.A  ==  1.A
2.CD ==  (1.C + 1.D)
2.E  ==  1.E
2.G  ==  1.G
2.H  ==  1.H
2.I  ==  1.I

D  Input visualization

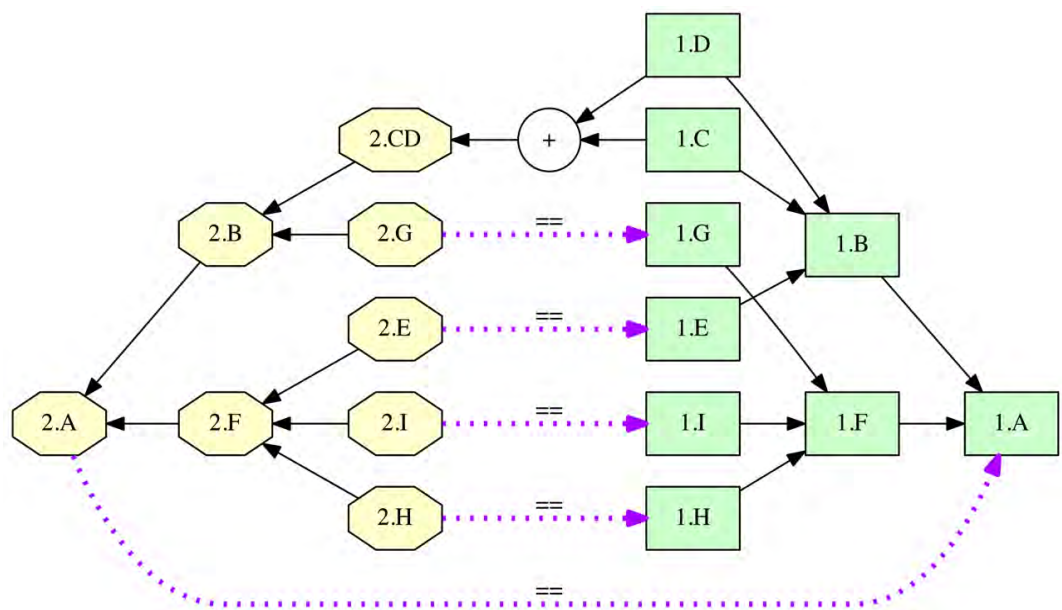

Merge legend (2E and 2F)

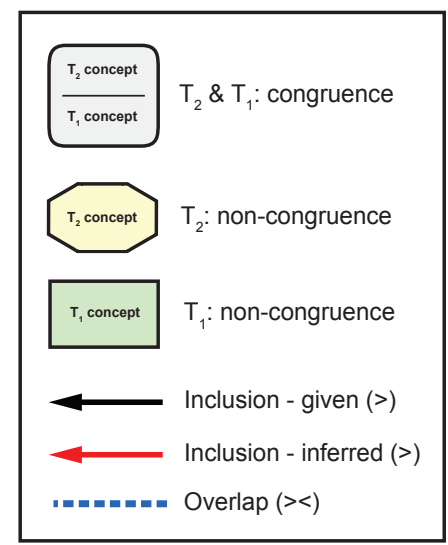

E  Containment (><) visualization

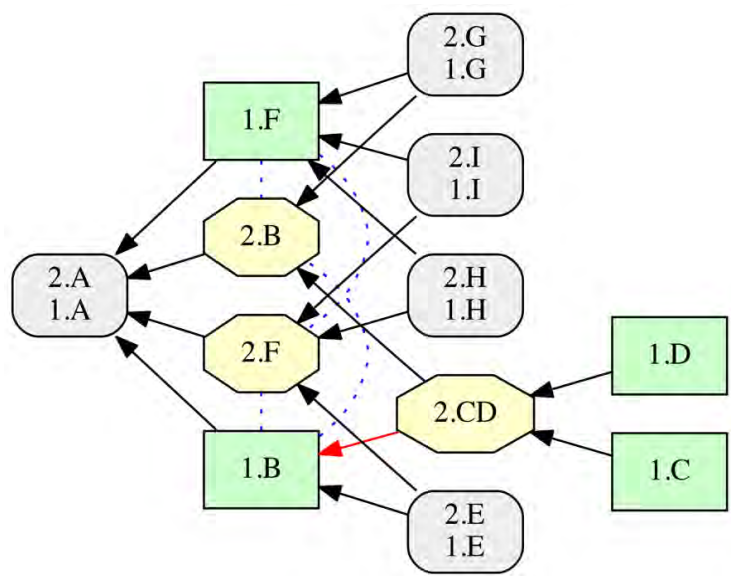

F  Merge concept visualization

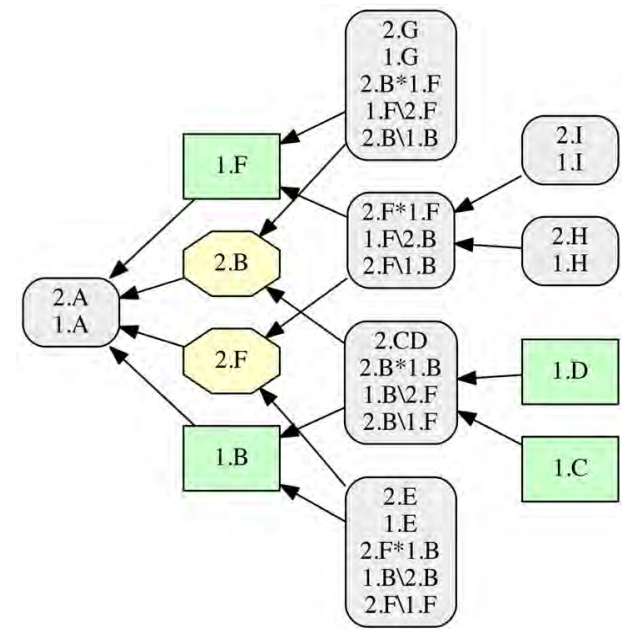

# Figure 3

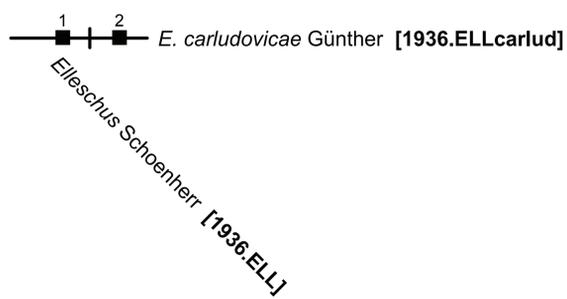
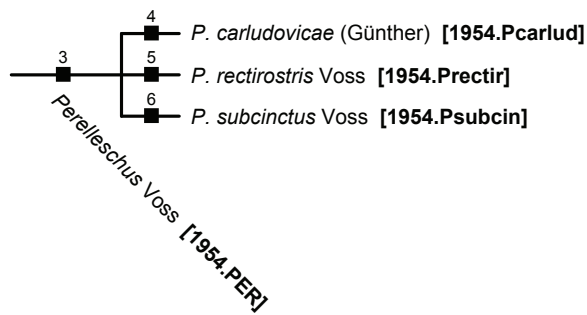
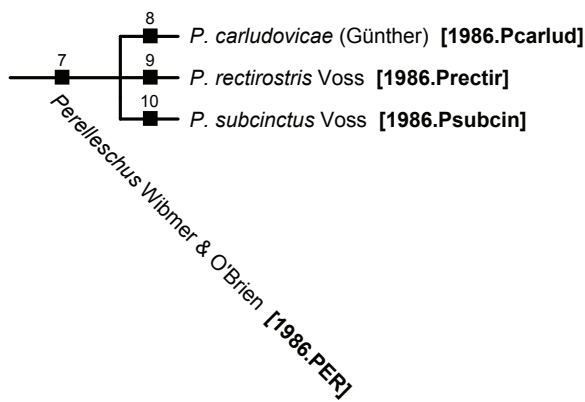
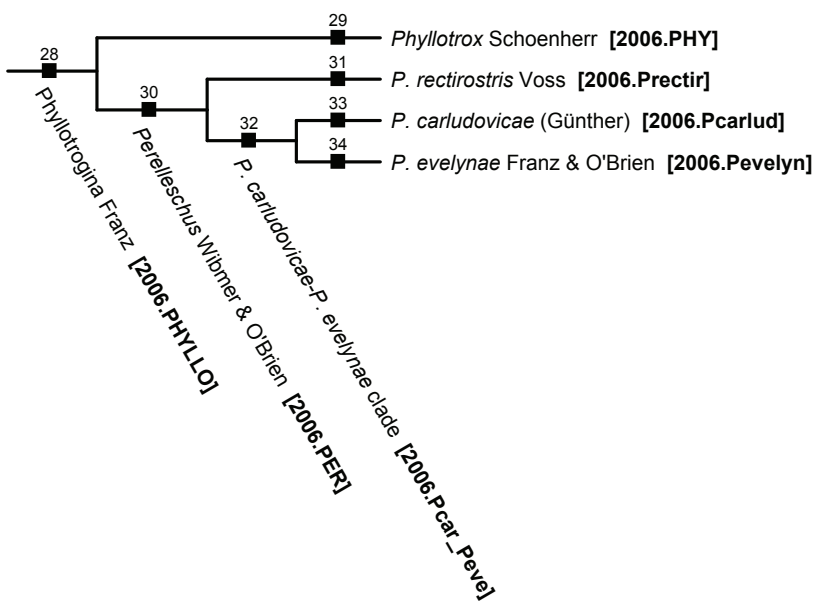
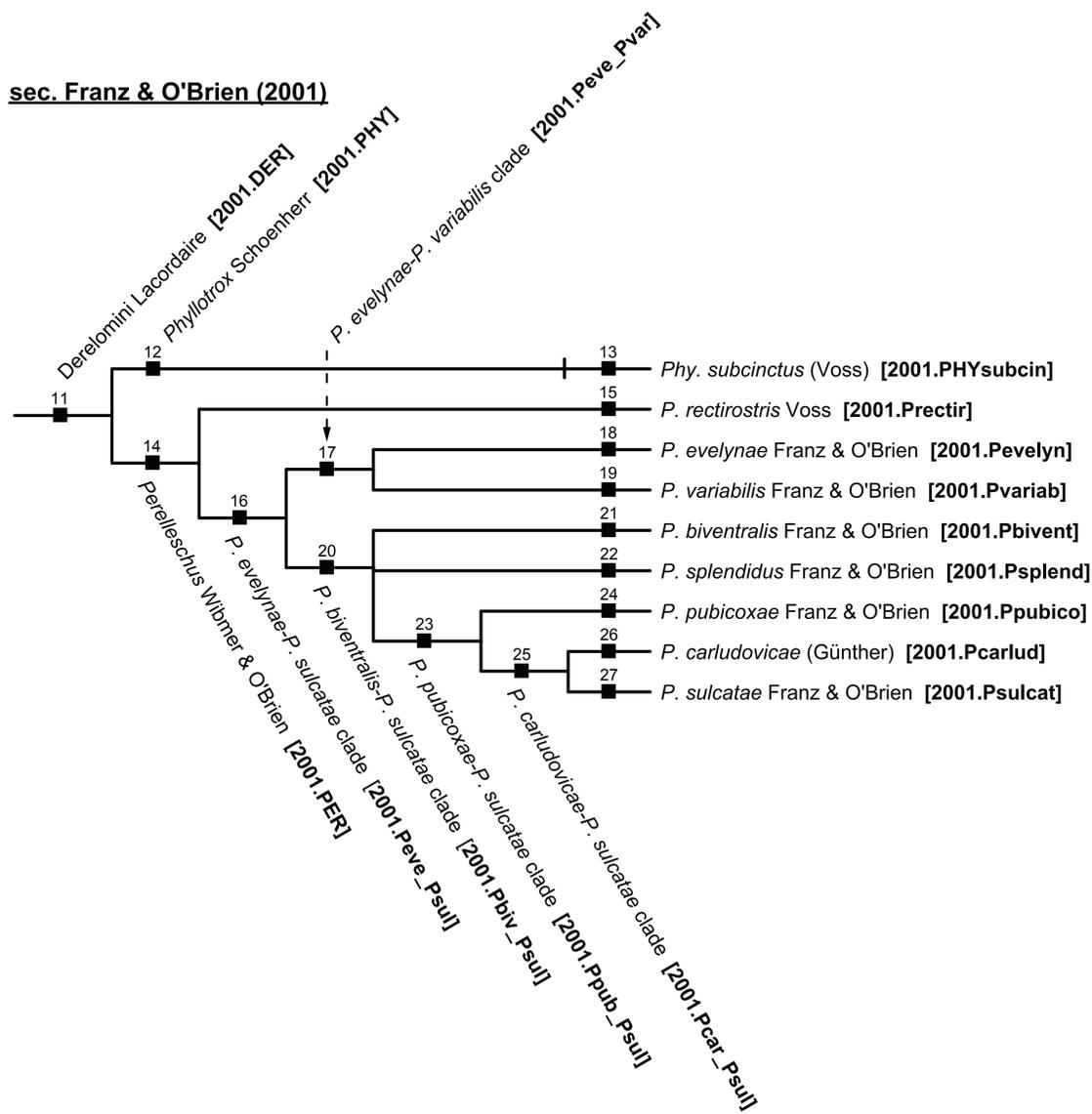
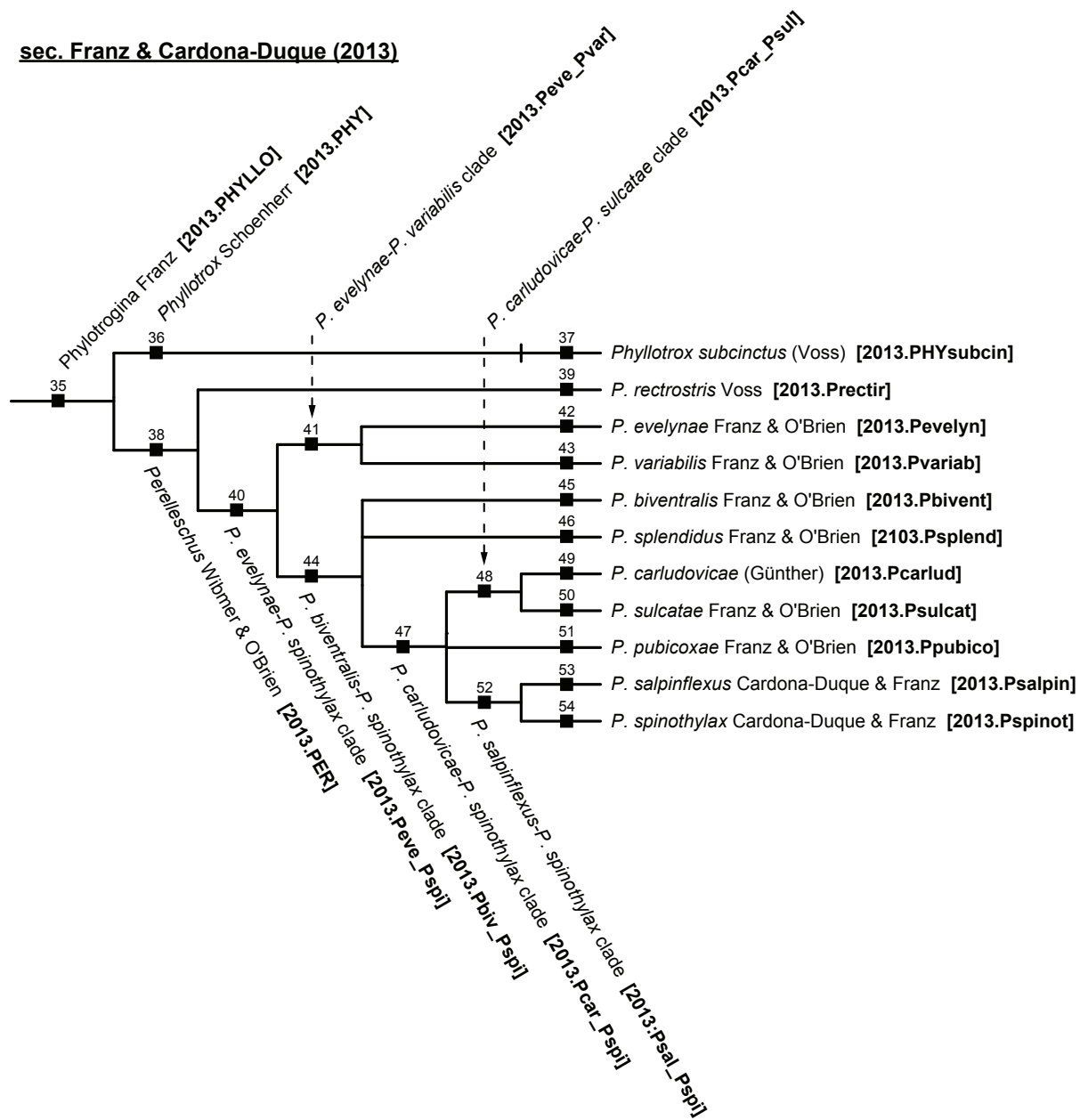

# Figure 4

**taxonomy 1954 Voss**
(PER Pcarlud Prectir Psubcin)

**taxonomy 1936 Günther**
(ELL ELLcarlud)

**articulation figure4**
1954.PER  > or < or ><  1936.ELL
1954.Pcarlud  ==  1936.ELLcarlud *
1954.Prectir  |  1936.ELLcarlud *
1954.Psubcin  |  1936.ELLcarlud *

**Output MIR - Deduced**
1954.PER  >  1936.ELLcarlud
1954.Pcarlud  ==  1936.ELLcarlud
1954.Prectir  |  1936.ELLcarlud
1954.Psubcin  |  1936.ELLcarlud

**Output MIR - Inferred**
1954.PER  >  1936.ELL
1954.Pcarlud  ==  1936.ELL
1954.Prectir  |  1936.ELL
1954.Psubcin  |  1936.ELL

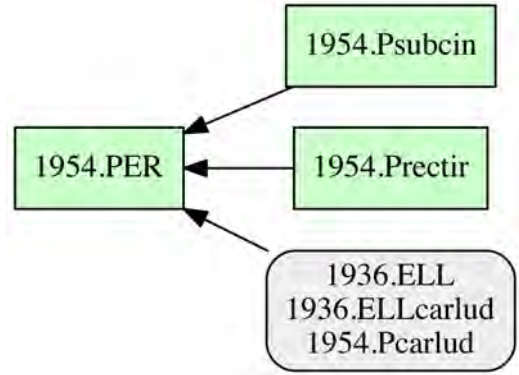

# Figure 5

**taxonomy 1954 Voss**
(PER Pcarlud Prectir Psubcin)

**taxonomy 1936 Günther**
(ELL ELL_IC ELLcarlud)

**articulation figure5**
**1954.PER >< 1936.ELL**
**1954.PER | 1936.ELL_IC** *
1954.Pcarlud == 1936.ELLcarlud *
1954.Prectir | 1936.ELLcarlud *
1954.Psubcin | 1936.ELLcarlud *

**Output MIR - Deduced**
1954.PER >< 1936.ELL
1954.PER | 1936.ELL_IC
1954.PER > 1936.ELLcarlud
1954.Pcarlud | 1936.ELL_IC
1954.Pcarlud == 1936.ELLcarlud
1954.Prectir | 1936.ELLcarlud
1954.Psubcin | 1936.ELLcarlud

**Output MIR - Inferred**
1954.Pcarlud < 1936.ELL
1954.Prectir | 1936.ELL
1954.Prectir | 1936.ELL_IC
1954.Psubcin | 1936.ELL
1954.Psubcin | 1936.ELL_IC

A 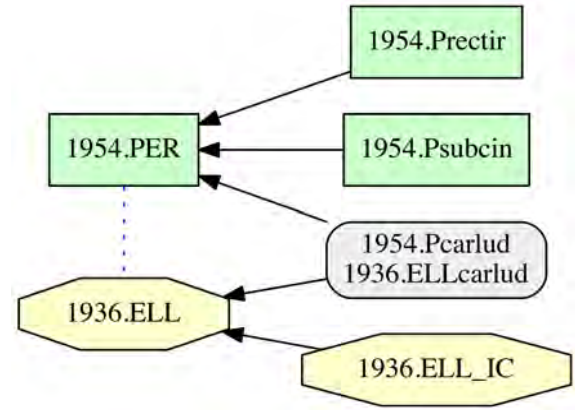

B 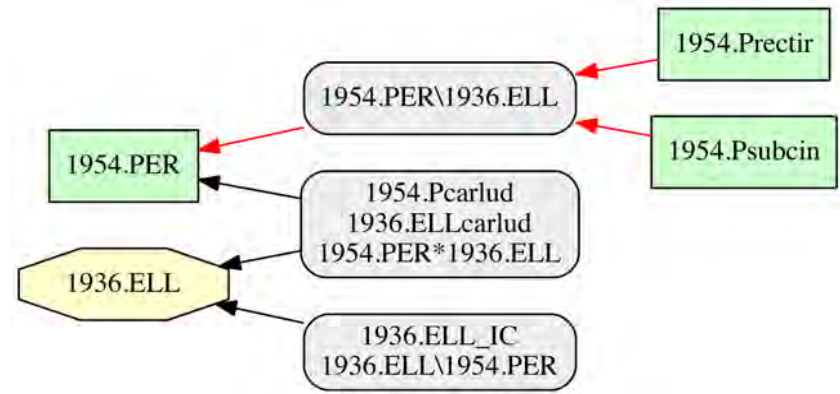

# Figure 6

**taxonomy 1954 Voss**
(PER Pcarlud Prectir Psubcin)

**taxonomy 1936 Günther**
(ELL ELL_**IC** ELLcarlud)

**articulation figure6**
1954.PER  <  1936.ELL
1954.Pcarlud  ==  1936.ELLcarlud
1954.Prectir  |  1936.ELLcarlud
1954.Psubcin  |  1936.ELLcarlud

**Output MIR - Deduced**
1954.PER  <  1936.ELL
1954.PER  >  1936.ELLcarlud
1954.Pcarlud  |  1936.ELL_**IC**
1954.Pcarlud  ==  1936.ELLcarlud
1954.Prectir  |  1936.ELLcarlud
1954.Psubcin  |  1936.ELLcarlud

**Output MIR - Inferred**
1954.PER  ><  1936.ELL_**IC**
1954.Pcarlud  <  1936.ELL
1954.Prectir  <  1936.ELL
1954.Prectir  <  1936.ELL_**IC**
1954.Psubcin  <  1936.ELL
1954.Psubcin  <  1936.ELL_**IC**

A 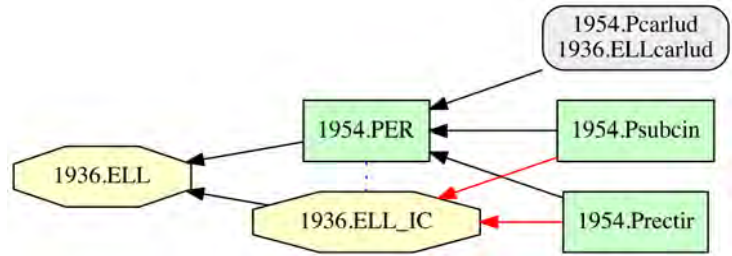

B 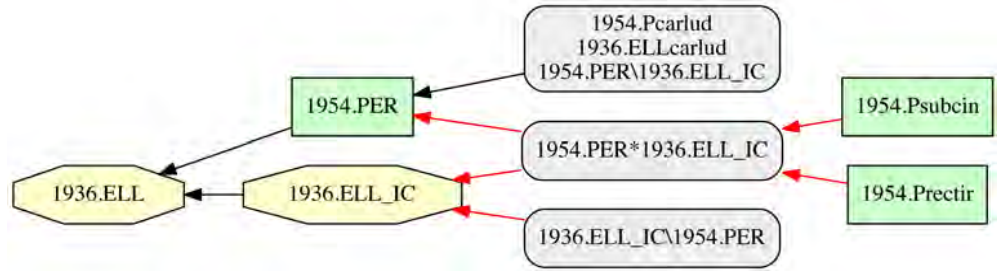

# Figure 7

**taxonomy 1954 Voss**
(PER Pcarlud Prectir Psubcin)

**taxonomy 1936 Günther**
(ELL ELL_IC ELLcarlud)

**articulation figure7**
**1954.PER >< 1936.ELL**
1954.Pcarlud == 1936.ELLcarlud
1954.Prectir | 1936.ELLcarlud
1954.Psubcin | 1936.ELLcarlud

**Output MIR - Deduced**
1954.PER >< 1936.ELL
1954.PER > 1936.ELLcarlud
1954.Pcarlud | 1936.ELL_IC
1954.Pcarlud == 1936.ELLcarlud
1954.Prectir | 1936.ELLcarlud
1954.Psubcin | 1936.ELLcarlud

**Output MIR - Inferred**
1954.PER >< or | 1936.ELL_IC
1954.Pcarlud < 1936.ELL
1954.Prectir < or >< or | 1936.ELL
1954.Prectir < or >< or | 1936.ELL_IC
1954.Psubcin < or >< or | 1936.ELL
1954.Psubcin < or >< or | 1936.ELL_IC

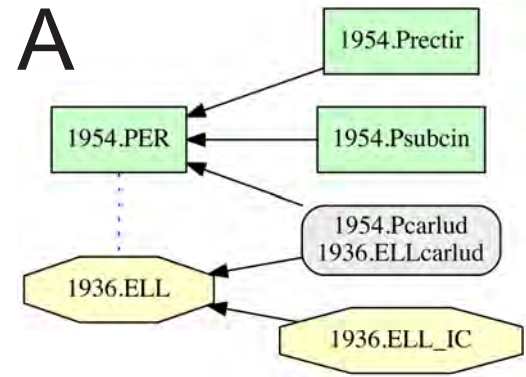
A
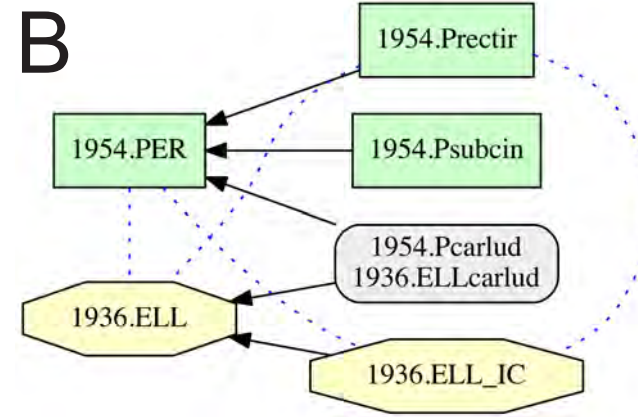
B
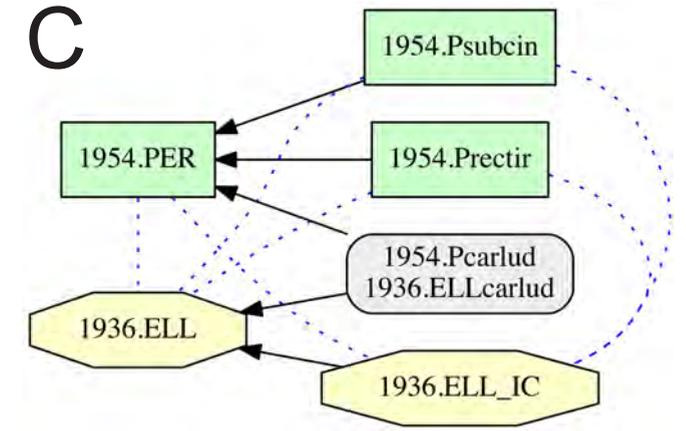
C
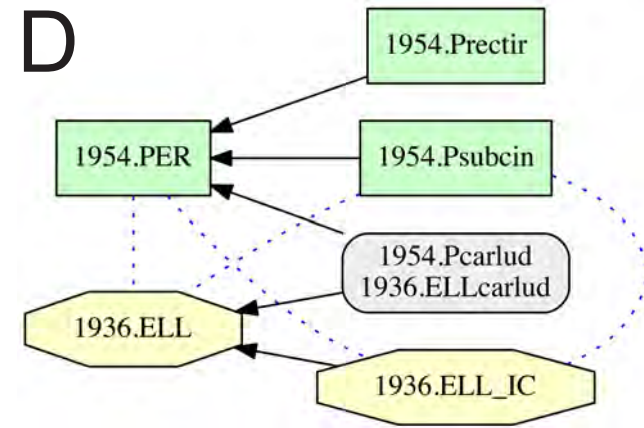
D
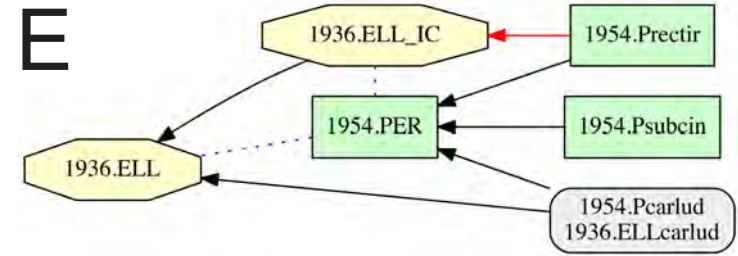
E
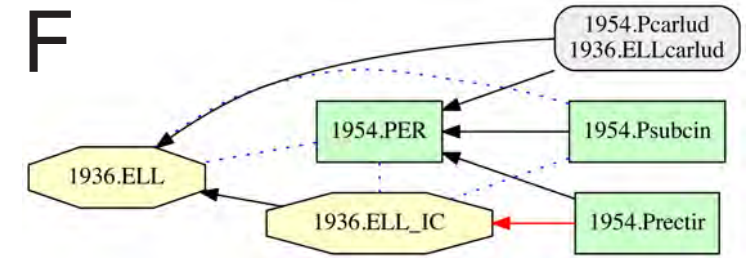
F
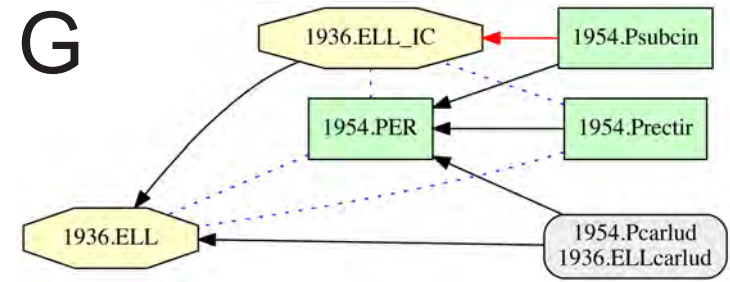
G
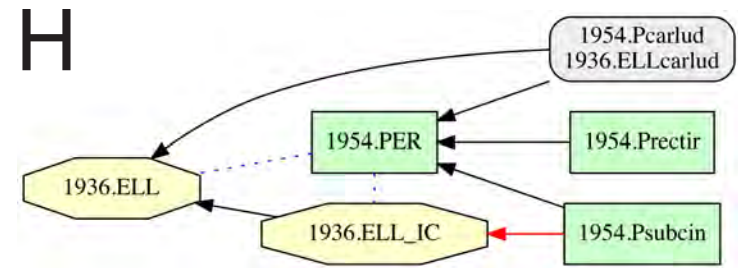
H

# Figure 8

**taxonomy 1954 Voss**
(PER Pcarlud Prectir Psubcin)

**taxonomy 1936 Günther**
(ELL ELL_IC ELLcarlud)

**articulation figure8**
1954.Pcarlud  ==  1936.ELLcarlud
1954.Prectir  |  1936.ELLcarlud
1954.Psubcin  |  1936.ELLcarlud

**Output MIR - Deduced**
1954.PER  >  1936.ELLcarlud
1954.Pcarlud  ==  1936.ELLcarlud
1954.Pcarlud  |  1936.ELL_IC
1954.Prectir  |  1936.ELLcarlud
1954.Psubcin  |  1936.ELLcarlud

**Output MIR - Inferred**
1954.PER  == or > or < or ><  1936.ELL
1954.PER  > or >< or |  1936.ELL_IC
1954.Pcarlud  <  1936.ELL
1954.Prectir  < or >< or |  1936.ELL
1954.Prectir  == > or < or >< or |  1936.ELL_IC
1954.Psubcin  < or >< or |  1936.ELL
1954.Psubcin  == or > or < or >< or |  1936.ELL_IC

**+ All 8 possible world merges of figure 7**

A 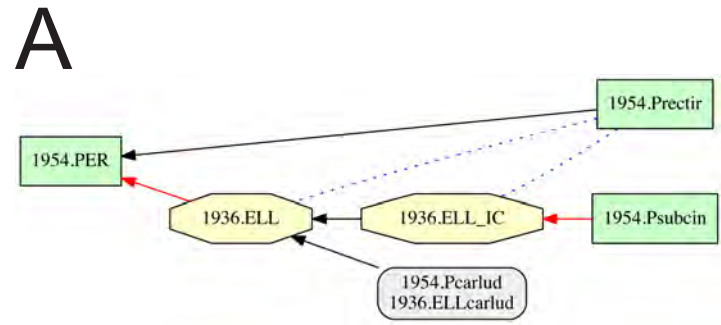

B 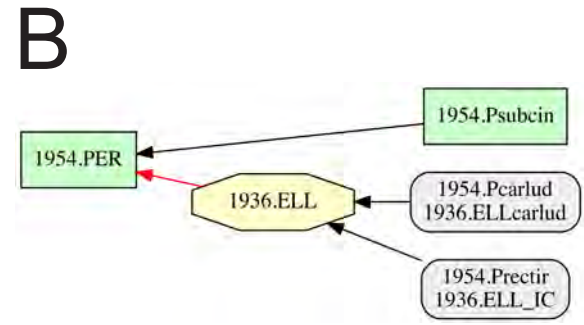

C 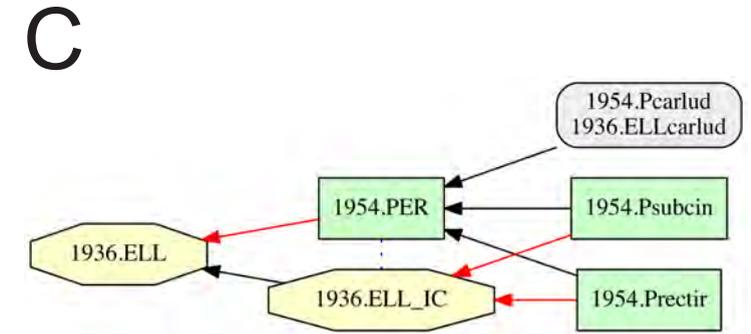

D 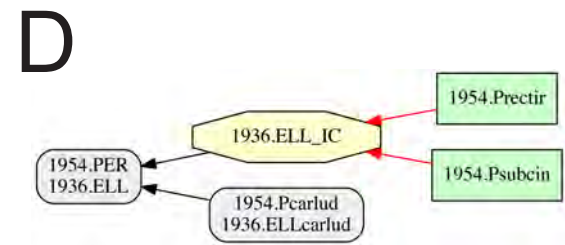

E 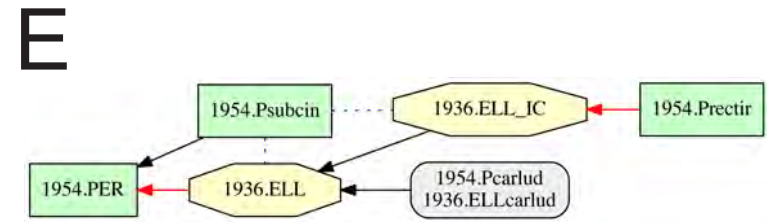

F 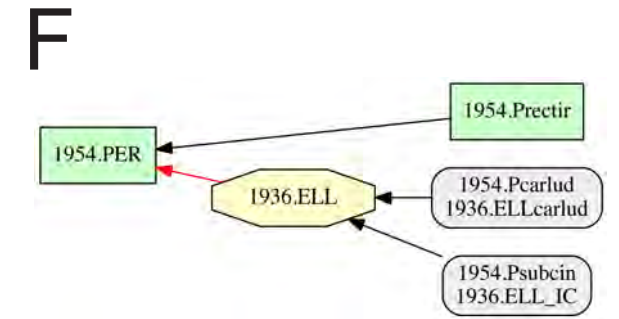

G 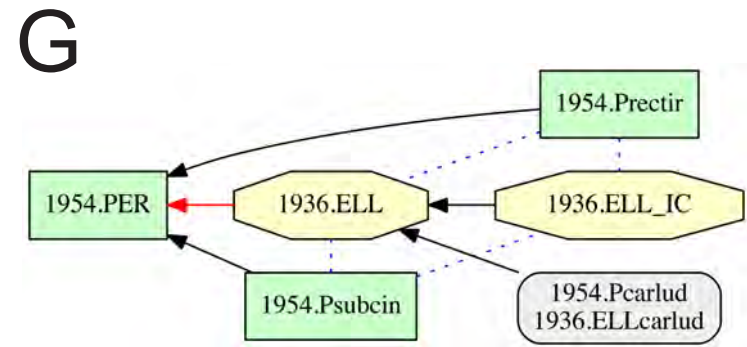

H 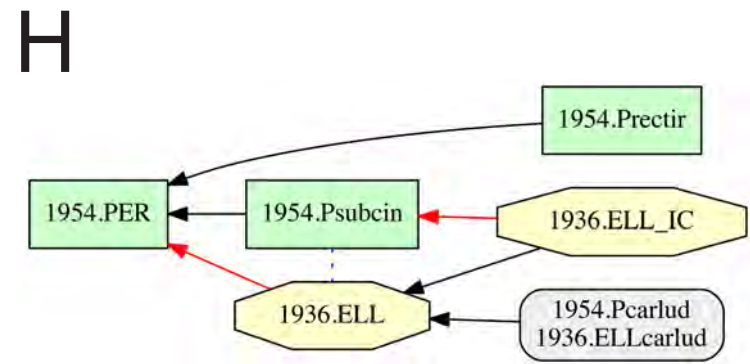

I 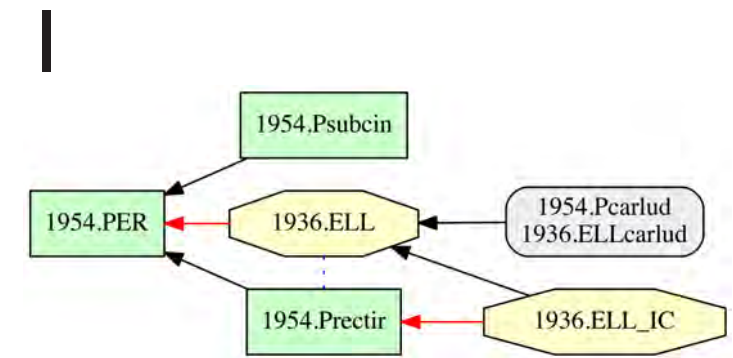

# Figure 9

**taxonomy 1986 Wibmer & O'Brien1986**
(PER Pcarlud Prectir Psubcin)

**taxonomy 1954 Voss**
(PER Pcarlud Prectir Psubcin)

**articulation figure9**
1986.PER == 1954.PER
1986.Pcarlud == 1954.Pcarlud *
1986.Prectir == 1954.Prectir *
1986.Psubcin == 1954.Psubcin *

| Nodes | |
|---|---|
| comb | 4 |
| **Edges** | |
| input | 3 |

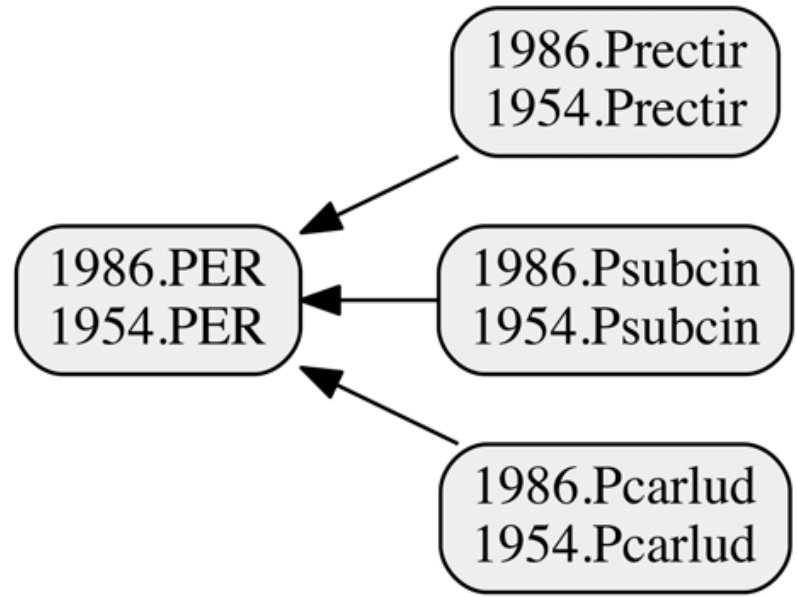

# Figure 10

**taxonomy 2001 Franz & O'Brien**
(DER PHY PER)
(PHY PHY_**IC** PHYsubcin)
(PER Prectir Peve_Psul)
(Peve_Psul Peve_Pvar Pbiv_Psul)
(Peve_Pvar Pevelyn Pvariab)
(Pbiv_Psul Pbivent Psplend Ppub_Psul)
(Ppub_Psul Ppubico Pcar_Psul)
(Pcar_Psul Pcarlud Psulcat)

**taxonomy 1986 Wibmer & O'Brien**
(PER Pcarlud Prectir Psubcin)

**articulation figure10**
2001.DER  >  1986.PER
2001.PHY  ><  1986.PER
2001.PHYsubcin  ==  1986.Psubcin *
2001.PER  ><  1986.PER
2001.Prectir  ==  1986.Prectir *
2001.Pcarlud  ==  1986.Pcarlud *

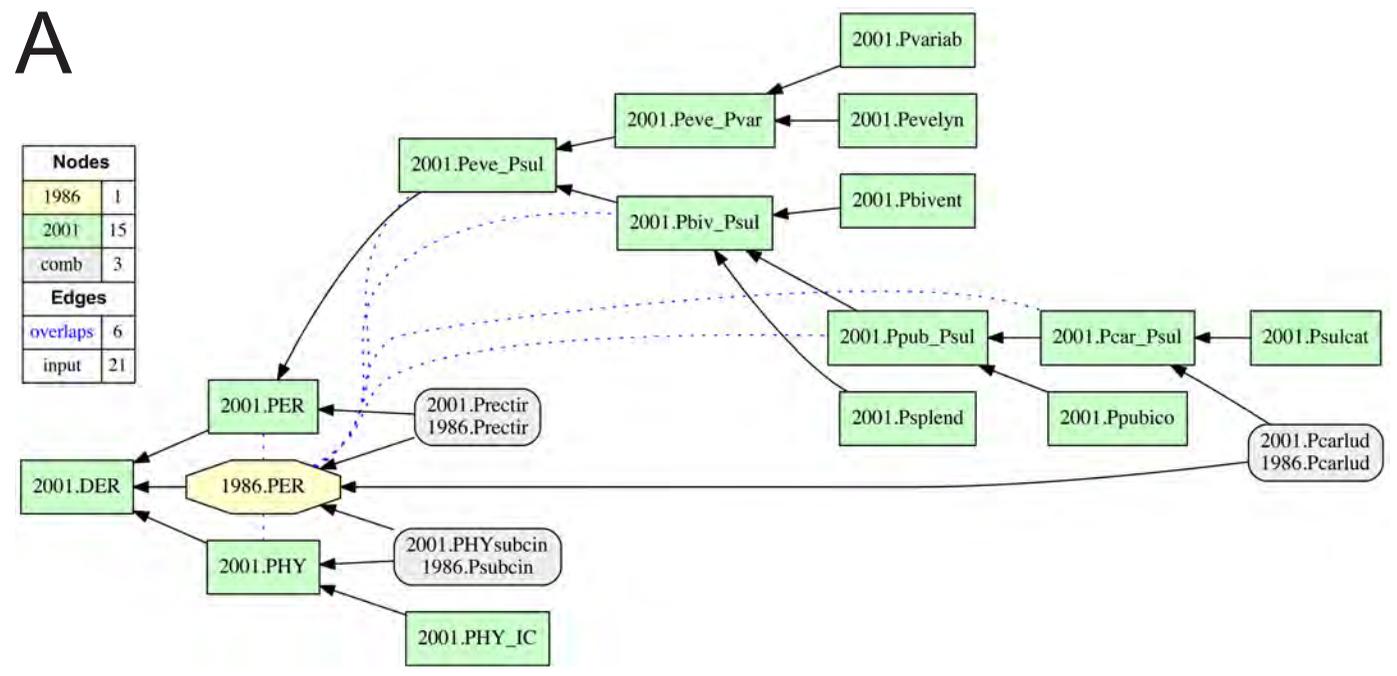

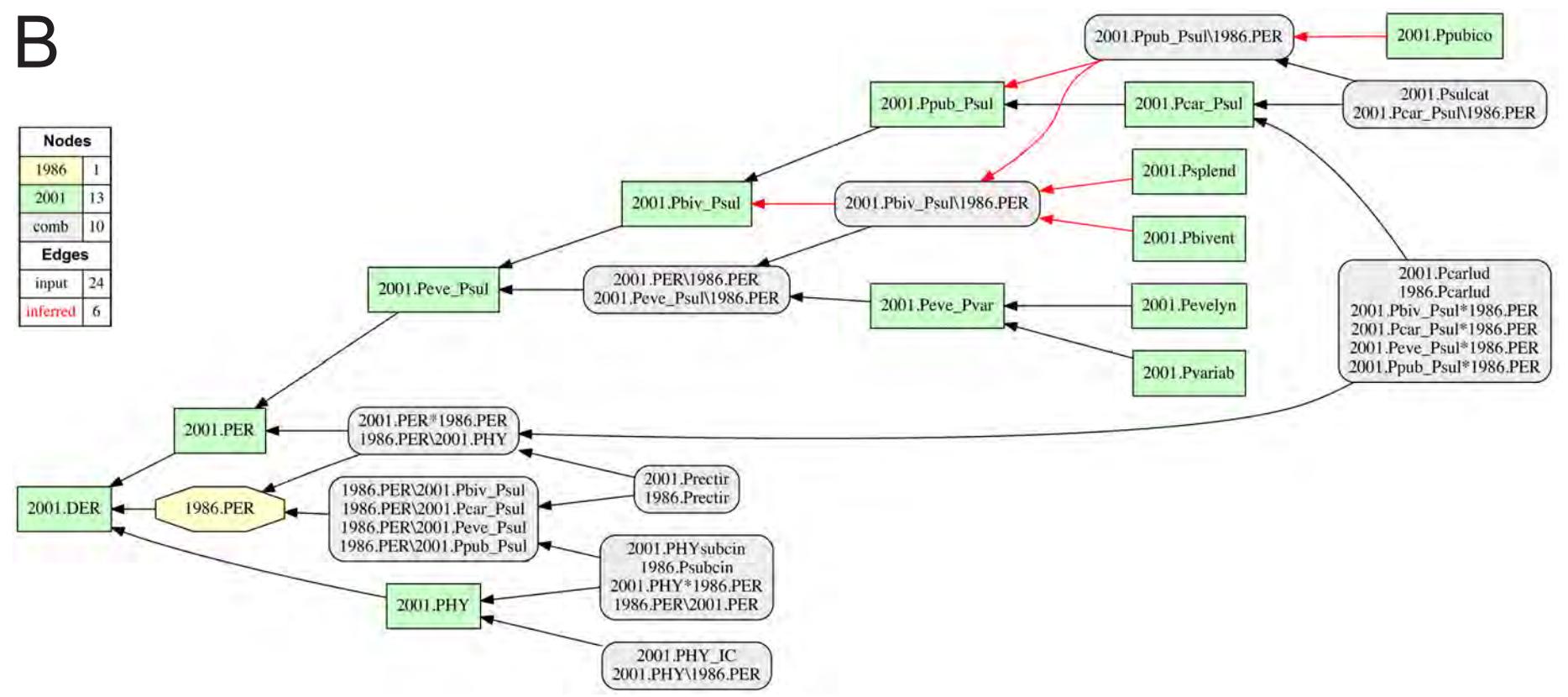

# Figure 11

**taxonomy 2006 Franz**
(PHYLLO PHY PER)
(PER Prectir Pcar_Peve)
(Pcar_Peve Pcarlud Pevelyn)

**taxonomy 2001 Franz & O'Brien**
(DER PHY PER)
(PHY PHYsubcin **nc**)
(PER Prectir Peve_Psul)
(Peve_Psul Peve_Pvar Pbiv_Psul)
(Peve_Pvar Pevelyn Pvariab)
(Pbiv_Psul Pbivent Psplend Ppub_Psul)
(Ppub_Psul Ppubico Pcar_Psul)
(Pcar_Psul Pcarlud Psulcat)

**articulation figure 11**
**2006.PHYLLO  <  2001.DER   [OST]**  (not ><)
**2006.PHY  ==  2001.PHY *** (not ><)
2006.PHY  >  2001.PHYsubcin
**2006.PER  <  2001.PER   [OST]**
2006.Prectir  ==  2001.Prectir *
**2006.Pcar_Peve  <  2001.Peve_Psul   [OST]**
**2006.Pcar_Peve  ><  2001.Peve_Pvar   [OST]**
**2006.Pcar_Peve  |  2001.Pvariab   [OST] ***
**2006.Pcar_Peve  ><  2001.Pbiv_Psul   [OST]**
**2006.Pcar_Peve  |  2001.Pbivent   [OST] ***
**2006.Pcar_Peve  |  2001.Psplend   [OST] ***
**2006.Pcar_Peve  ><  2001.Ppub_Psul   [OST]**
**2006.Pcar_Peve  |  2001.Ppubico   [OST] ***
**2006.Pcar_Peve  ><  2001.Pcar_Psul   [OST]**
**2006.Pcar_Peve  |  2001.Psulcat   [OST] ***
2006.Pcarlud  ==  2001.Pcarlud *
2006.Pevelyn  ==  2001.Pevelyn *

A
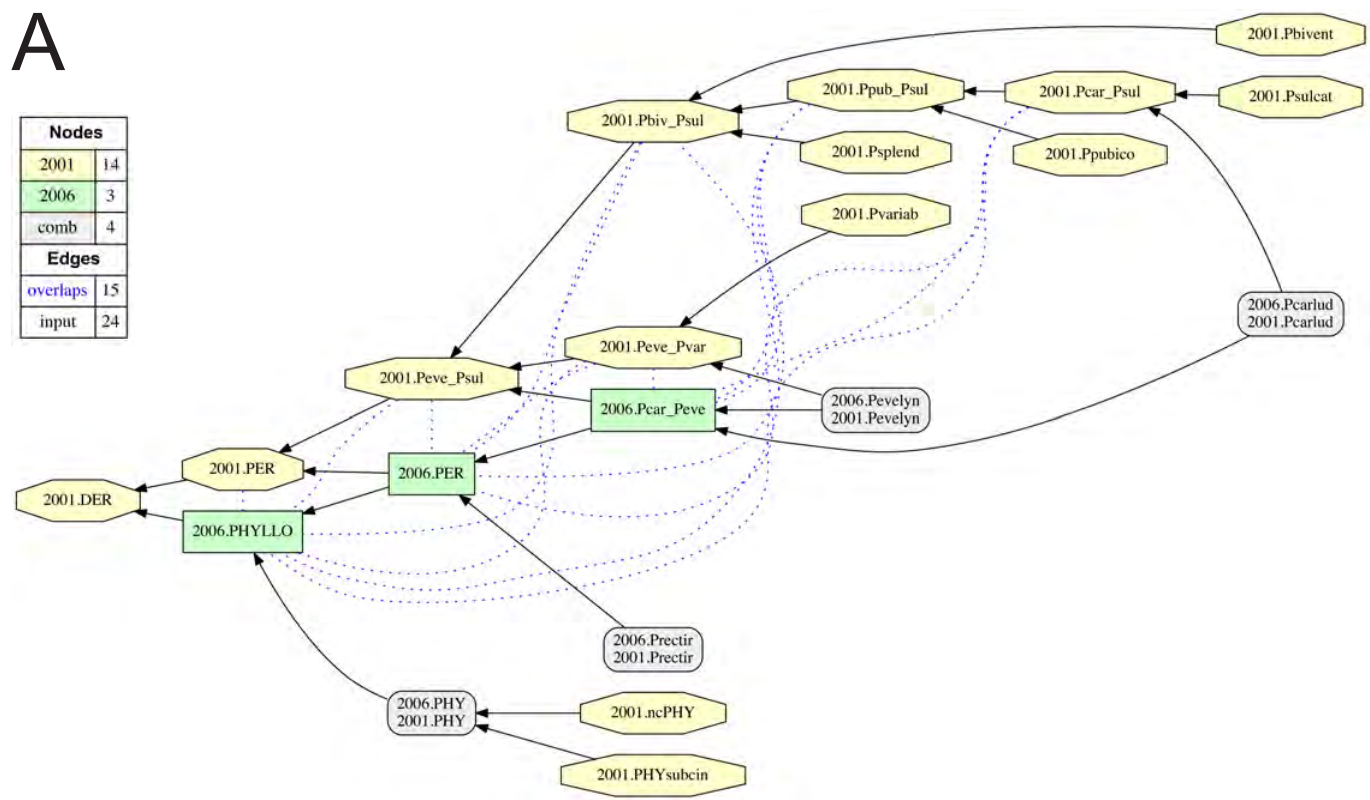

B
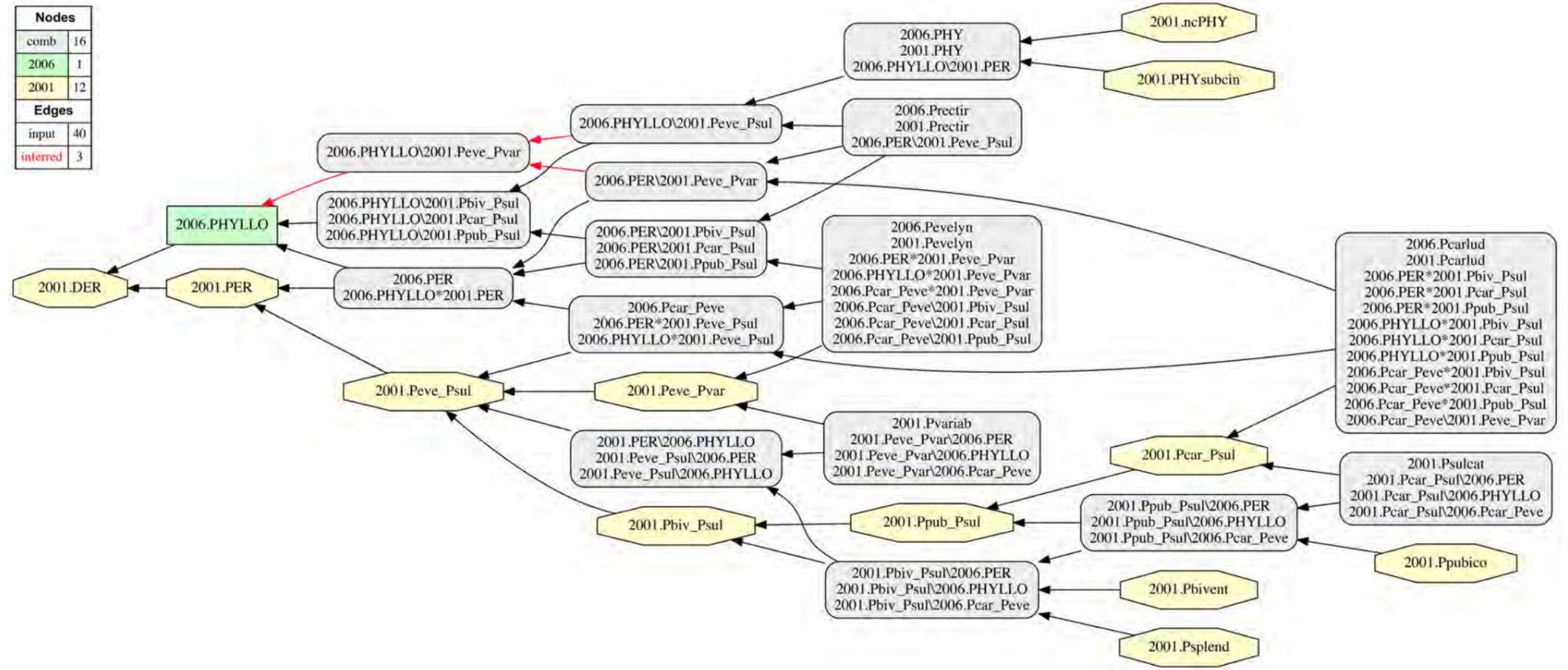

# Figure 12

**taxonomy 2006 Franz**
(PHYLLO PHY PER)
(PER Prectir Pcar_Peve)
(Pcar_Peve Pcar_Peve_**IC** Pcarlud Pevelyn)

**taxonomy 2001 Franz & O'Brien**
(DER PHY PER)
(PHY PHYsubcin **nc**)
(PER Prectir Peve_Psul)
(Peve_Psul Peve_Pvar Pbiv_Psul)
(Peve_Pvar Pevelyn Pvariab)
(Pbiv_Psul Pbivent Psplend Ppub_Psul)
(Ppu_bPsul Ppubico Pcar_Psul)
(Pcar_Psul Pcarlud Psulcat)

**articulation figure12**
**2006.PHYLLO == 2001.DER  [INT] ***   (not ><)
**2006.PHY == 2001.PHY ***   (not ><)
2006.PHY > 2001.PHYsubcin
**2006.PER == 2001.PER  [INT]**
2006.Prectir == 2001.Prectir *
**2006.Pcar_Peve == 2001.Peve_Psul  [INT]**
**2006.Pcar_Peve > 2001.Peve_Pvar  [INT]**
**2006.Pcar_Peve_IC > 2001.Pvariab  [INT] ***
**2006.Pcar_Peve > 2001.Pbiv_Psul  [INT]**
**2006.Pcar_Peve_IC > 2001.Pbivent  [INT] ***
**2006.Pcar_Peve_IC > 2001.Psplend  [INT] ***
**2006.Pcar_Peve > 2001.Ppub_Psul  [INT]**
**2006.Pcar_Peve_IC > 2001.Ppubico  [INT] ***
**2006.Pcar_Peve > 2001.Pcar_Psul  [INT]**
**2006.Pcar_Peve_IC > 2001.Psulcat  [INT] ***
2006.Pcarlud == 2001.Pcarlud *
2006.Pevelyn == 2001.Pevelyn *

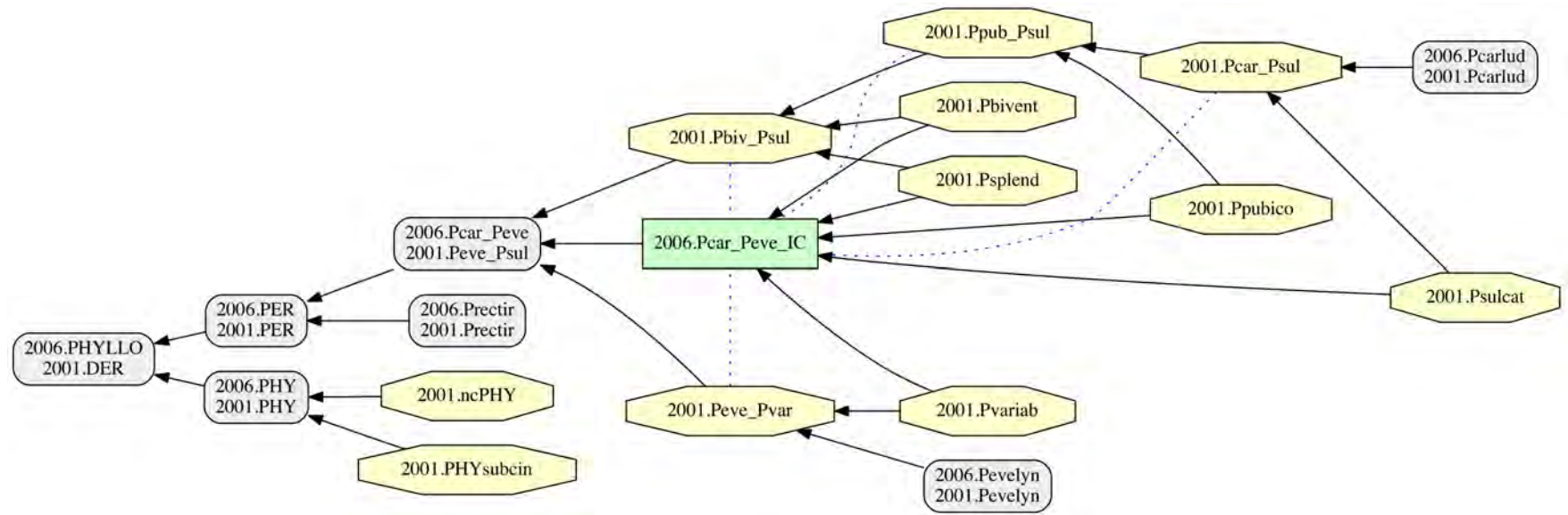

A

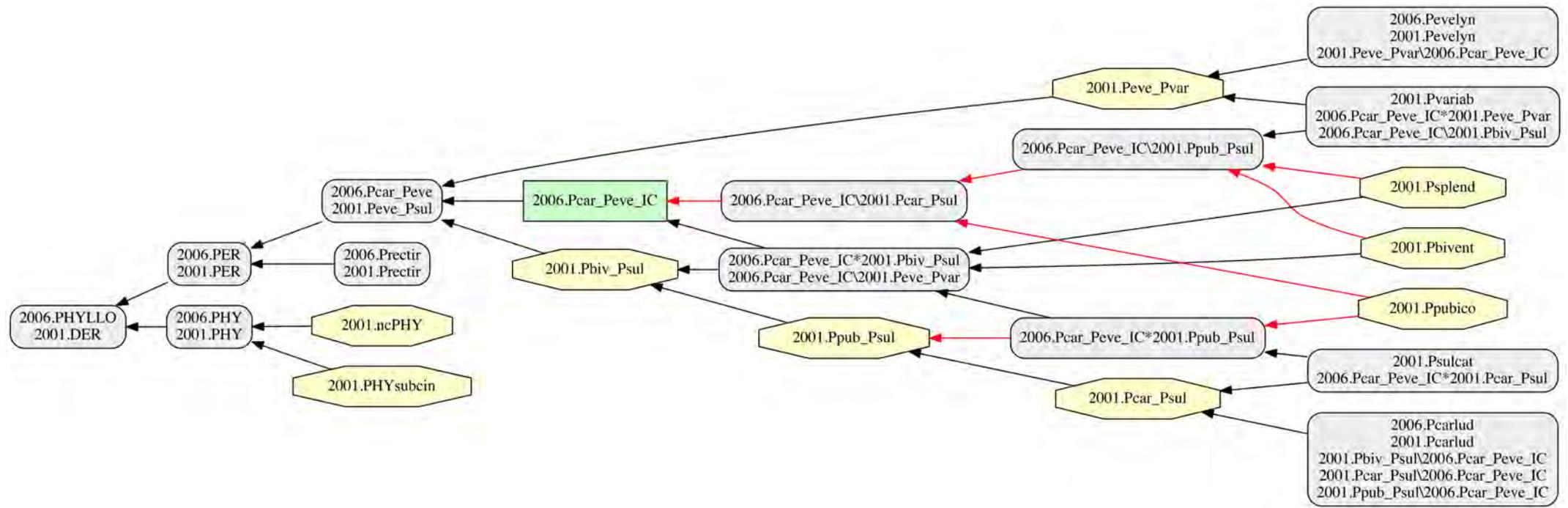

B

# Figure 13

**taxonomy 2013 Franz & Cardona-Duque**
(PHYLLO PHY PER)
(PHY PHYsubcin **nc**)
(PER Prectir Peve_Pspi)
(Peve_Pspi Peve_Pvar Pbiv_Pspi)
(Peve_Pvar Pevelyn Pvariab)
(Pbiv_Pspi Pbivent Psplend Pcar_Pspi)
(Pcar_Pspi Pcar_Psul Ppubico Psal_Pspi)
(Pcar_Psul Pcarlud Psulcat)
(Psal_Pspi Psalpin Pspinot)

**taxonomy 2006 Franz**
(PHYLLO PHY PER)
(PER Prectir Pcar_Peve)
(Pcar_Peve Pcarlud Pevelyn)

**articulation figure13**
**2013.PHYLLO > 2006.PHYLLO  [OST]**   (not ==)
2013.PHY ==  2006.PHY *
2013.PHYsubcin  <  2006.PHY
**2013.PER  >  2006.PER  [OST]**
2013.Prectir ==  2006.Prectir *
**2013.Peve_Pspi  ><  2006.PER   [OST]**
**2013.Peve_Pspi  >  2006.Pcar_Peve   [OST]**
**2013.Peve_Pvar  ><  2006.PER   [OST]**
**2013.Peve_Pvar  ><  2006.Pcar_Peve   [OST]**
2013.Pevelyn  ==  2006.Pevelyn *
**2013.Pvariab  |  2006.Pcar_Peve   [OST] ***
**2013.Pbiv_Pspi  ><  2006.PER   [OST]**
**2013.Pbiv_Pspi  ><  2006.Pcar_Peve   [OST]**
**2013.Pbivent  |  2006.Pcar_Peve   [OST] ***
**2013.Psplend  |  2006.Pcar_Peve   [OST] ***
**2013.Pcar_Pspi  ><  2006.Pcar_Peve   [OST]**
**2013.Pcar_Psul  ><  2006.Pcar_Peve   [OST]**
2013.Pcarlud  ==  2006.Pcarlud *
**2013.Psulcat  |  2006.Pcar_Peve   [OST] ***
**2013.Ppubico  |  2006.Pcar_Peve   [OST] ***
**2013.Psal_Pspi  |  2006.Pcar_Peve   [OST]**
**2013.Psalpin  |  2006.Pcar_Peve   [OST] ***
**2013.Pspinot  |  2006.Pcar_Peve   [OST] ***

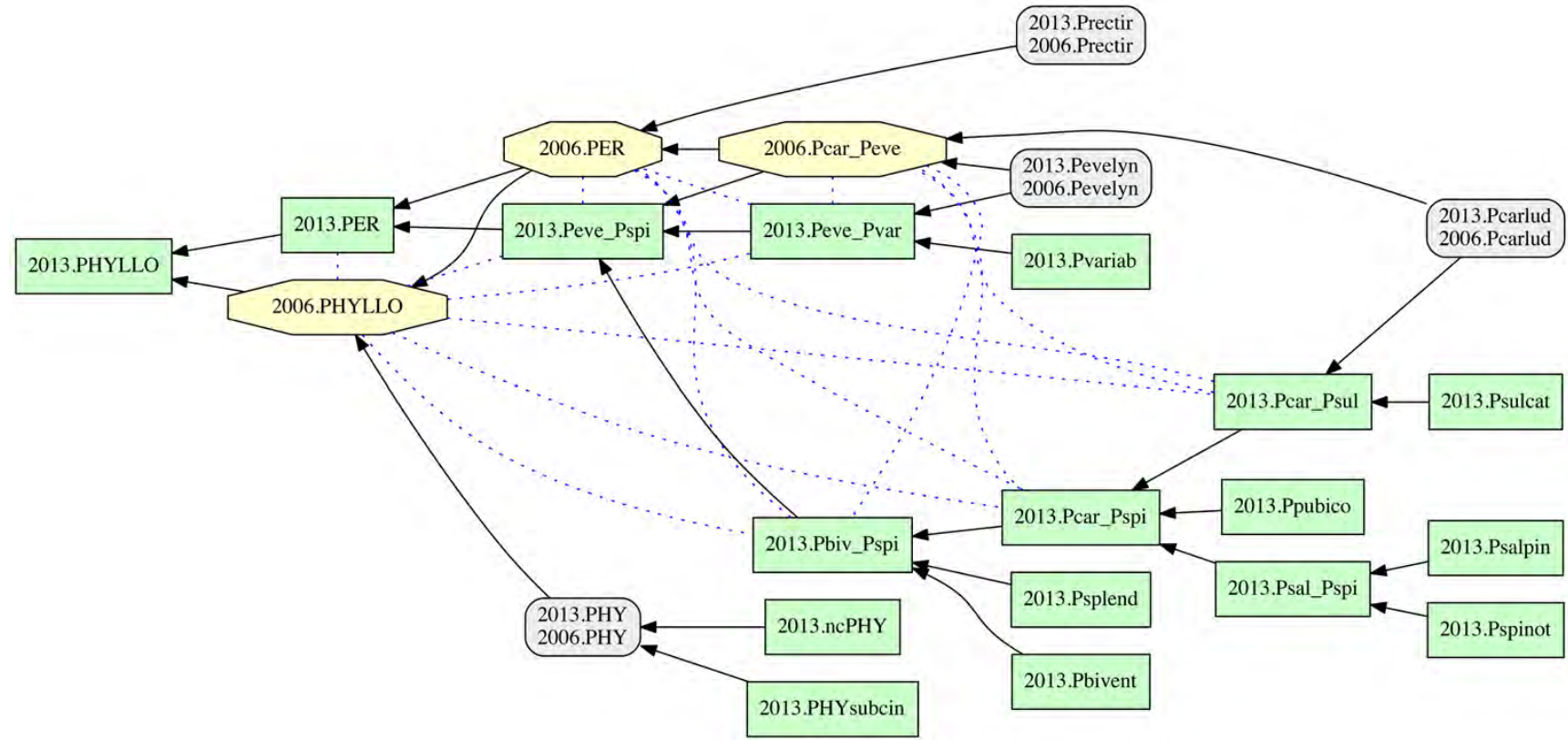

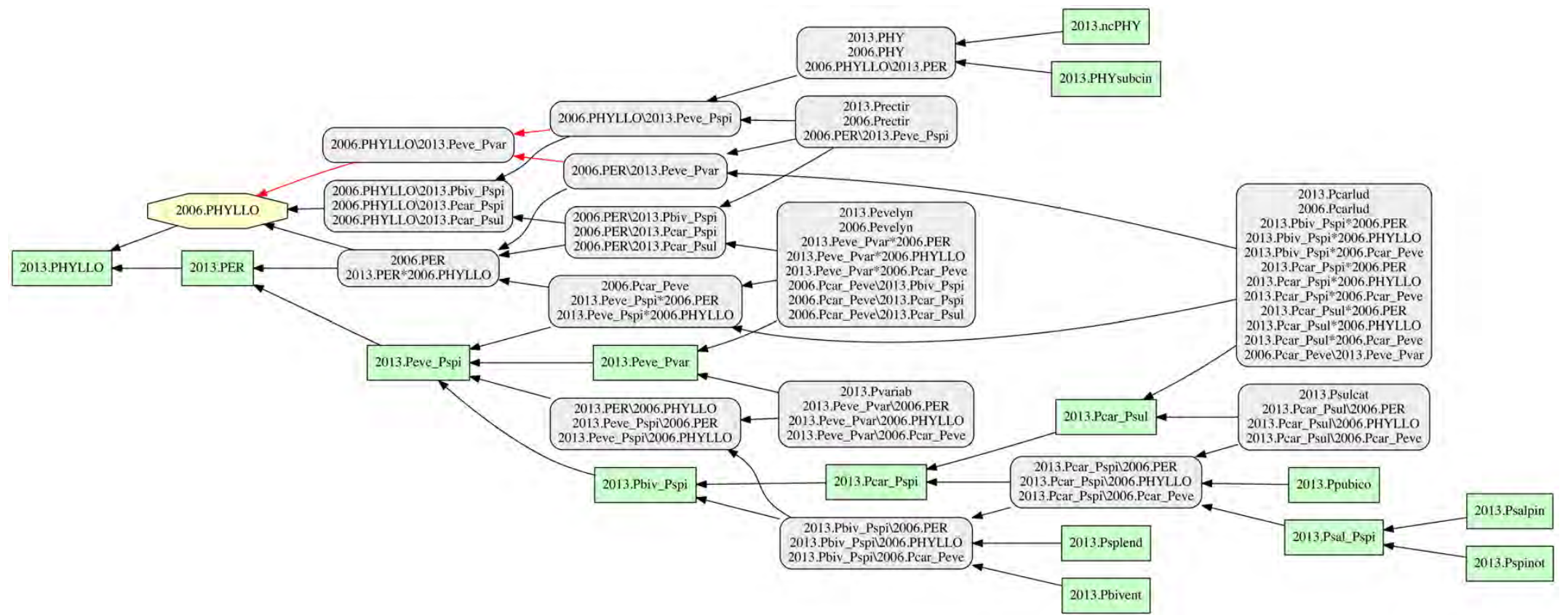

# Figure 14

**taxonomy 2013 Franz & Cardona-Duque**
(PHYLLO PHY PER)
(PHY PHYsubcin **nc**)
(PER Prectir Peve_Pspi)
(Peve_Pspi Peve_Pvar Pbiv_Pspi)
(Peve_Pvar Pevelyn Pvariab)
(Pbiv_Pspi Pbivent Psplend Pcar_Pspi)
(Pcar_Pspi Pcar_Psul Ppubico Psal_Pspi)
(Pcar_Psul Pcarlud Psulcat)
(Psal_Pspi Psalpin Pspinot)

**taxonomy 2006 Franz**
(PHYLLO PHY PER)
(PER Prectir Pcar_Peve)
(Pcar_Peve Pcar_Peve_**IC** Pcarlud Pevelyn)

**articulation figure14**
**2013.PHYLLO == 2006.PHYLLO [INT]** *
2013.PHY == 2006.PHY *
2013.PHYsubcin < 2006.PHY
**2013.PER == 2006.PER [INT]**
2013.Prectir == 2006.Prectir *
**2013.Peve_Pspi < 2006.PER [INT]**
**2013.Peve_Pspi == 2006.Pcar_Peve [INT]** * (not >)
**2013.Peve_Pvar < 2006.PER [INT]**
**2013.Peve_Pvar < 2006.Pcar_Peve [INT]**
2013.Pevelyn == 2006.Pevelyn *
**2013.Pvariab < 2006.Pcar_Peve_IC [INT]** *
**2013.Pbiv_Pspi < 2006.PER [INT]**
**2013.Pbiv_Pspi < 2006.Pcar_Peve [INT]**
**2013.Pbivent < 2006.Pcar_Peve_IC [INT]** *
**2013.Psplend < 2006.Pcar_Peve_IC [INT]** *
**2013.Pcar_Pspi < 2006.Pcar_Peve [INT]**
**2013.Pcar_Psul < 2006.Pcar_Peve [INT]**
2013.Pcarlud == 2006.Pcarlud *
**2013.Psulcat < 2006.Pcar_Peve_IC [INT]** *
**2013.Ppubico < 2006.Pcar_Peve_IC [INT]** *
**2013.Psal_Pspi < 2006.Pcar_Peve [INT]**
**2013.Psalpin < 2006.Pcar_Peve_IC [INT]** *
**2013.Pspinot < 2006.Pcar_Peve_IC [INT]** *

# Figure 15

**taxonomy 2013 Franz & Cardona-Duque**
(PHYLLO PHY PER)
(PHY PHYsubcin)
(PER Prectir Peve_Pspi)
(Peve_Pspi Peve_Pvar Pbiv_Pspi)
(Peve_Pvar Pevelyn Pvariab)
(Pbiv_Pspi Pbivent Psplend Pcar_Pspi)
(Pcar_Pspi Pcar_Psul Ppubico Psal_Pspi)
(Pcar_Psul Pcarlud Psulcat)
(Psal_Pspi Psalpin Pspinot)

**taxonomy 2001 Franz & O'Brien**
(DER PHY PER)
(PHY PHYsubcin)
(PER Prectir Peve_Psul)
(Peve_Psul Peve_Pvar Pbiv_Psul)
(Peve_Pvar Pevelyn Pvariab)
(Pbiv_Psul Pbivent Psplend Ppub_Psul)
(Ppub_Psul Ppubico Pcar_Psul)
(Pcar_Psul Pcarlud Psulcat)

**articulation figure15**
**2013.PHYLLO > 2001.DER  [OST]**  (not ><)
2013.PHY  ==  2001.PHY   (not ><)
2013.PHYsubcin  ==  2001.PHYsubcin *
**2013.PER  >  2001.PER  [OST]**
2013.Prectir  ==  2001.Prectir *
**2013.Peve_Pspi  >  2001.Peve_Psul   [OST]**
2013.Peve_Pvar  ==  2001.Peve_Pvar
2013.Pevelyn  ==  2001.Pevelyn *
2013.Pvariab  ==  2001.Pvariab *
**2013.Pbiv_Pspi  >  2001.Pbiv_Psul   [OST]**
2013.Pbivent  ==  2001.Pbivent *
2013.Psplend  ==  2001.Psplend *
**2013.Pcar_Pspi  >  2001.Ppub_Psul   [OST]**
2013.Pcar_Psul  ==  2001.Pcar_Psul
2013.Pcarlud  ==  2001.Pcarlud *
2013.Psulcat  ==  2001.Psulcat *
2013.Ppubico  ==  2001.Ppubico *
**2013.Psal_Pspi  |  2001.Ppub_Psul   [OST] ***
**2013.Psalpin  |  2001.Ppub_Psul   [OST] ***
**2013.Pspinot  |  2001.Ppub_Psul   [OST] ***

A

B

# Figure 16

**taxonomy 2013 Franz & Cardona-Duque**
(PHYLLO PHY PER)
(PHY PHYsubcin)
(PER Prectir Peve_Pspi)
(Peve_Pspi Peve_Pvar Pbiv_Pspi)
(Peve_Pvar Pevelyn Pvariab)
(Pbiv_Pspi Pbivent Psplend Pcar_Pspi)
(Pcar_Pspi Pcar_Psul Ppubico Psal_Pspi)
(Pcar_Psul Pcarlud Psulcat)
(Psal_Pspi Psalpin Pspinot)

**taxonomy 2001 Franz & O'Brien**
(DER PHY PER)
(PHY PHYsubcin)
(PER Prectir Peve_Psul)
(Peve_Psul Peve_Pvar Pbiv_Psul)
(Peve_Pvar Pevelyn Pvariab)
(Pbiv_Psul Pbivent Psplend Ppub_Psul)
(Ppub_Psul Ppub_Psul_IC Ppubico Pcar_Psul)
(Pcar_Psul Pcarlud Psulcat)

**articulation figure16**
2013.PHYLLO == 2001.DER   [INT] *  (not ><)
**2013.PHY == 2001.PHY**  (not ><)
2013.PHYsubcin == 2001.PHYsubcin *
**2013.PER == 2001.PER    [INT]**
2013.Prectir == 2001.Prectir *
**2013.Peve_Pspi == 2001.Peve_Psul   [INT]**
2013.Peve_Pvar == 2001.Peve_Pvar
2013.Pevelyn == 2001.Pevelyn *
2013.Pvariab == 2001.Pvariab *
**2013.Pbiv_Pspi == 2001.Pbiv_Psul   [INT]**
2013.Pbivent == 2001.Pbivent *
2013.Psplend == 2001.Psplend *
**2013.Pcar_Pspi == 2001.Ppub_Psul   [INT]**
2013.Pcar_Psul == 2001.Pcar_Psul
2013.Pcarlud == 2001.Pcarlud *
2013.Psulcat == 2001.Psulcat *
2013.Ppubico == 2001.Ppubico *
**2013.Psal_Pspi < 2001.Ppub_Psul   [INT]**
**2013.Psalpin < 2001.Ppub_Psul_IC   [INT] ***
**2013.Pspinot < 2001.Ppub_Psul_IC   [INT] ***

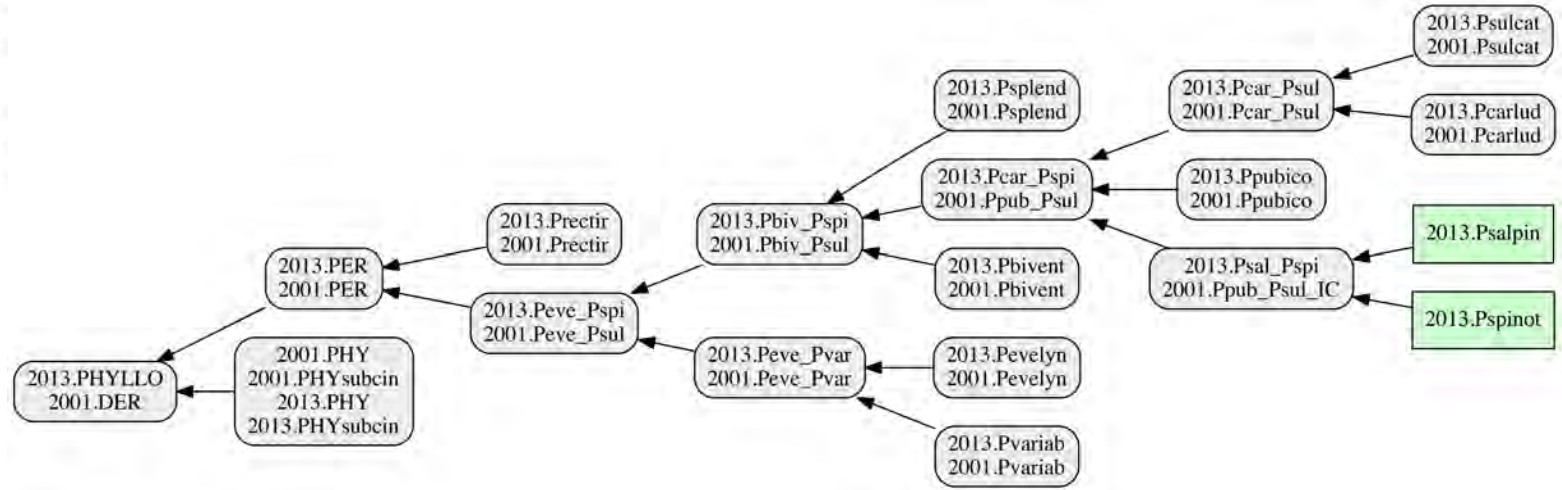